\documentclass[paper]{JHEP3}
\usepackage{axodraw}
\newcommand{\bfig}{\begin{center}\begin{picture}}
\newcommand{\efig}[1]{\end{picture}\\{\small #1}\end{center}}

\newcommand{\bmip}[2]{\begin{minipage}[t]{#1pt}\bfig(#1,#2)}
\newcommand{\emip}[1]{\efig{#1}\end{minipage}}

\newcommand{\bq}{\begin{equation}}
\newcommand{\eq}{\end{equation}}
\newcommand{\bqa}{\begin{eqnarray}}
\newcommand{\eqa}{\end{eqnarray}}

\newcommand{\benon}{\begin{equation*}}
\newcommand{\eenon}{\end{equation*}}
\newcommand{\beanon}{\begin{eqnarray*}}
\newcommand{\eeanon}{\end{eqnarray*}}

\newcommand{\vsk}{\vskip 10 pt\noindent}
\newcommand{\hsk}{\hskip 10 pt\noindent}
\newcommand{\inp}[2]{\vsk\underline{{\tt\large{#1}}, ({\tt{#2}})}: }
\newcommand{\inpe}[1]{\vsk\underline{{\tt\large{#1}}}: }
\newcommand{\pha}{{\tt PHASE }}
\newcommand{\phanosp}{{\tt PHASE}}

\newcommand{\LHP}{Les Houches Protocol}

\newcommand{\Parctan}{{\mathrm{arctan}}}
\newcommand{\rd}{{\mathrm{d}}}
\newcommand{\Pd}{\mathrm d}

\newcommand{\Pf}{\mathrm f}

\newcommand{\Pl}{\mathrm l}
\newcommand{\Pp}{\mathrm p}
\newcommand{\Pq}{\mathrm q}
\newcommand{\Pu}{\mathrm u}

\newcommand{\Ps}{\mathrm s}
\newcommand{\Pb}{\mathrm b}
\newcommand{\Pc}{\mathrm c}
\newcommand{\Pt}{\mathrm t}
\newcommand{\PW}{\mathrm W}
\newcommand{\PZ}{\mathrm Z}
\newcommand{\PH}{\mathrm H}
\newcommand{\PV}{\mathrm V}
\newcommand{\PM}{\mathrm M}
\newcommand{\PE}{\mathrm E}
\newcommand{\PT}{\mathrm P_T}
\newcommand{\GeV}{\mathrm{GeV}}

\newcommand{\MZ}{\mathrm {M_\PZ}}
\newcommand{\MW}{\mathrm {M_\PW}}
\newcommand{\Mt}{\mathrm {M_\Pt}}
\newcommand{\Mb}{\mathrm {M_\Pb}}

\newcommand{\ol}{\overline}
\newcommand{\bea}{\begin{eqnarray}}
\newcommand{\eea}{\end{eqnarray}}

        \newdimen\mysep                
        \newdimen\hmysep
\def    \be             {\begin{equation}}
\def    \ee             {\end{equation}}
\def    \ba             {\begin{eqnarray}}
\def    \ea             {\end{eqnarray}}

\def    \=              {\;=\;}
\def    \frac           #1#2{{#1 \over #2}}

\def\refeq#1{\mbox{(\ref{#1})}}

\def    \rd             {{\mathrm d}}    
  
\def    \bra#1          {\mbox{$\langle #1 |$}}
\def    \ket#1          {\mbox{$| #1 \rangle$}}

\newcommand     \MSB            {\ifmmode {\overline{\rm MS}} \else 
                                 $\overline{\rm MS}$  \fi}

\def    \as             {\ifmmode \alpha_s \else $\alpha_s$ \fi}

\def\rt1{\raisebox{-1ex}{\rlap{$\; \rho \to 1 \;\;$}}
\raisebox{.4ex}{$\;\; \;\;\simeq \;\;\;\;$}}

\def\si{\sigma}

\def\met{$\rlap{\kern.2em/}E_T$}

\def\refse#1{\mbox{Sect.~\ref{#1}}}

\def\pl #1 #2 #3 {{\it Phys.~Lett.} {\bf#1} (#2) #3}   
\def\np #1 #2 #3 {{\it Nucl.~Phys.} {\bf#1} (#2) #3}
\def\zp #1 #2 #3 {{\it Z.~Phys.} {\bf#1} (#2) #3}
\def\pr #1 #2 #3 {{\it Phys.~Rev.} {\bf#1} (#2) #3}
\def\prep #1 #2 #3 {{\it Phys.~Rep.} {\bf#1} (#2) #3}
\def\prl #1 #2 #3 {{\it Phys.~Rev.~Lett.} {\bf#1} (#2) #3}
\def\intj #1 #2 #3 {{\it Int. J. Mod. Phys.} {\bf#1} (#2) #3}
\def\mpl #1 #2 #3 {{\it Mod.~Phys.~Lett.} {\bf#1} (#2) #3}
\def\rmp #1 #2 #3 {{\it Rev. Mod. Phys.} {\bf#1} (#2) #3}
\def\cpc #1 #2 #3 {{\it Comp. Phys. Commun.} {\bf#1} (#2) #3}
\def\epj #1 #2 #3 {{\it Eur. Phys. J.} {\bf#1} (#2) #3}

\title{ 
      PHASE, a Monte Carlo event generator for 
      six-fermion physics at the LHC.
      \thanks{Work supported by the Ministero dell'Istruzione,
      dell'Universit\`a e della Ricerca under contract 2004021808\_009.
      The work of EA is supported by the 
      MIUR under Contract ``Rientro dei cervelli'' Decreto MIUR 26-01-2001 
      N.13 .}}
\vfill                                                       
\author{
  Elena ACCOMANDO, Alessandro BALLESTRERO
  and Ezio MAINA\\
Dipartimento di Fisica Teorica, Universit\`{a} di Torino and INFN Sezione di 
Torino, 
\\ Via P. Giuria 1, 10125-Torino, Italy
\\ E-mail: \email{accomand@to.infn.it, ballestr@to.infn.it,
 maina@to.infn.it}}
\abstract{\pha is a new event generator dedicated to the study of Standard 
  Model processes with six fermions in the final state at the LHC. The code is 
  intended for analyses of vector boson scattering, Higgs search, 
  three gauge boson production, and top physics. This first 
  version of the program describes final states characterized by the presence 
  of one neutrino, $\Pp\Pp\to 4\Pq +\Pl\nu_{\Pl}$, at \cal{O}($\alpha^6$). 
  \pha is based on a new 
  iterative-adaptive multichannel technique, and employs exact leading order 
  matrix elements. 
  The code can generate unweighted events for any subset of all
  available final states. The produced parton-level events carry full 
  information on their colour and flavour structure, enabling the evolution of 
  the partons into fully hadronised final states. An interface to 
  hadronization packages is provided via the \LHP.}

\preprint{DFTT 09/2005}
\begin{document}
\section{Introduction}
\label{sec:intro}

The large energies available in the forthcoming Large Hadron Collider (LHC)
will make it possible to access many-particle final states with much more 
statistics than before. Among these final states, six-fermion signals are of 
particular interest for several topics. They have a great potential in 
Higgs boson discovery and for analyzing vector boson scattering. The 
origin of the electroweak symmetry breaking is still an open problem. The most 
direct way to address this question is searching for the Higgs boson. At the 
LHC, the SM Higgs production is driven by gluon-gluon fusion. The fusion of 
$\PW$ and $\PZ$ gauge bosons ($\Pq\Pq\rightarrow\Pq\Pq\PH$) represents 
the second most important contribution to the Higgs production cross section. 
Among all possible final states which might be generated by this process, the 
Higgs decay channel into $\PW\PW$, giving rise to two forward-backward jets 
plus four leptons or two leptons and two jets from the $\PW$'s, is 
particularly clean. This channel has been found to be quite promising for the 
Higgs search in the low-intermediate mass range 
(115$\lesssim\PM_\PH\lesssim$200 GeV) 
favoured by present electroweak precision measurements (see for instance 
ref. \cite{higgs_vbf}). If the Higgs exists, 
kinematical configurations with six fermions in the final state are then an
important tool for its detection and for measuring its properties. If the 
Higgs is not present, the complementary approach to the question of 
electroweak symmetry breaking is studying vector boson scattering. In the 
absence of the Higgs, general arguments based on unitarity imply that massive 
gauge bosons 
become strongly interacting at the TeV scale. Processes mediated by massive 
vector boson scattering, $\PV\PV\rightarrow\PV\PV$ ($\PV$=$\PW$,$\PZ$), are 
then the most sensitive to the symmetry breaking mechanism. LHC will be able 
to produce for the first time processes containing boson-boson interactions at 
TeV scale ($\Pq\Pq\rightarrow\Pq\Pq\PV\PV\rightarrow 6\Pf$), offering a unique 
possibility to understand the nature of electroweak symmetry breaking. 
Six-fermion processes are also strictly related to the production of three 
vector bosons, which would allow to extract new informations on quartic 
self-couplings. Moreover, they open the window on the broad field of top quark 
physics. These reactions give in fact access to $\Pt\bar\Pt$ and single-top 
production in six-fermions, enabling measurements of top mass, $\PW\Pt\Pb$ 
coupling, decay branching ratios, rare decays and all other interesting 
features related to the top quark.
Finally, we should mention that multi-particle final states of this kind 
constitute a direct background to most searches for new physics. 
\par\noindent
This paper presents a new event generator, \pha \cite{phase}, which is 
designed to evaluate all Standard Model processes $\Pp\Pp\rightarrow 6\Pf$ in 
lowest order. The code is therefore particularly appropriate to compute and 
analyse Higgs physics, vector boson scattering and triple gauge boson 
production. This first version takes into account only quark/antiquark
initiated processes. Enabling the code to compute $\Pt\bar\Pt$ production, 
which receives its dominant contribution from gluon-gluon initiated processes, 
is one of the most important evolutions planned for the near future. We have 
built an event generator 
{\it dedicated} to all classes and topologies of final states specific for 
these studies. A recent example of {\it dedicated} program for LHC physics is 
{\tt Alpgen} \cite{Alpgen}. The complementary approach is 
given by {\it multi-purpose} programs for the automatic generation of any 
user-specified parton level process. The following codes for multi-parton 
production are available: {\tt Amegic-Sherpa} \cite{Amegic}, {\tt CompHEP} 
\cite{Comphep}, {\tt Grace-Gr@ppa} \cite{Grace}, {\tt Madevent} 
\cite{Madgraph}, {\tt Phegas \& Helac} \cite{Phegas}, 
{\tt O'Mega $\&$ Whizard} \cite{Omega}.
\par\noindent
In the following sections we give a full description and documentation of 
\phanosp.
Three are the key features of our code. The first one consists in the use of a 
modular helicity formalism for computing matrix elements. Scattering 
amplitudes get contributions from thousands of diagrams. In this context,  
computation efficiency has a primary role. The helicity method \cite{helamp}
we use is 
suited to compute in a fast and compact way parts of diagrams of increasing 
size, and recombine them later to obtain the final set. In this manner, parts 
common to various diagrams are evaluated just once for all possible helicity 
configurations, optimizing the computation 
procedure. The second main feature concerns the integration. We have devised a 
new integration method to address the crucial point of reaching good stability 
and efficiency in event generation. Our integration strategy combines the 
commonly used multichannel approach \cite{multichannel} with the adaptivity of 
{\tt VEGAS} \cite{vegas}. As the number of particles increases, the 
multichannel technique becomes rather cumbersome, given the thousands of 
resonant structures which can appear in the amplitude at the same time. For 
this reason, its efficiency in event generation is still debated. 
Conversely, {\tt VEGAS} adaptivity is not powerful enough to deal with 
all possible peaks of the amplitude. 
%The limiting key being the finite number of bins one can devide the 
%integration interval of each variable.
We have merged the two strategies in a single procedure. The outcome is that 
{\tt PHASE} adapts to different kinematical cuts and peaks with good 
efficiency, using only few channels per process. As third main feature, \pha 
employs the {\it one-shot} method developed for {\tt WPHACT} \cite{wphact}, 
and used for four-fermion data analyses at LEP2. In this running 
mode, all processes (of order 1000) are simultaneously generated in the 
correct relative proportion for any set of experimental cuts, and directly 
interfaced to hadronization and detector simulation programs, giving a fully 
comprehensive physical description. This possibility is relevant at the LHC, 
where one has to deal with a huge multiplicity of final states as well as 
initial states.
%which gives rise to final configurations which are not experimentally 
%distinguishable.

\refse{sec:structure} reviews the general structure of the code. 
\refse{sec:classes} provides a classification of the processes 
$\Pp\Pp\rightarrow 4\Pq+\Pl\nu_{\Pl}$.
\refse{sec:matrixelements} describes how matrix elements are computed.
\refse{sec:phasespace} explains the integration method, and 
\refse{sec:eventgeneration} addresses the event generation strategy, 
covering also aspects of shower evolution and hadronization. The two available
running modes of \pha are discussed in \refse{sec:modes}. 
Some applications of the code for Higgs boson production are presented in 
\refse{sec:sample}. Two technical 
appendixes describe in detail input parameters and {\it input-files} 
the user must provide. A summary is given in \refse{sec:conclusions}.

\section{General features of the program}
\label{sec:structure}

\pha is composed of several building blocks. A main body encloses the overall 
structure of the program, defining the sequence of operations via a set of 
subroutine calls. There are two possible running modes: {\it single-process} 
and {\it one-shot}. The sequence of operations depends on the selected mode. 
In the former case, the set of subroutine calls includes initialization of the 
selected process, and evaluation of cross section, integrand maxima and 
phase-space grids. The outcome obtained in {\it single-process} mode 
constitutes the essential ingredient of the {\it one-shot} run. In this latter 
mode, one generates unweighted events for all desired processes in the same 
run. In the following sections, we describe the general framework and criteria 
the sequence of main operations is based on.

\begin{table*}[htb]
\label{table:1}
\newcommand{\m}{\hphantom{$-$}}
\newcommand{\cc}[1]{\multicolumn{1}{c}{#1}}
\renewcommand{\tabcolsep}{2pc} % enlarge column spacing
\renewcommand{\arraystretch}{1.2} % enlarge line spacing
\begin{tabular}{@{}llll}
\hline
particles & type & diagrams & process number \\
\hline
$c\bar s d\bar u c\bar s l\bar\nu$ & 4$\PW$ & 202={\bf{101}}$\times$2 & 8\\
$u\bar u u\bar u c\bar s l\bar\nu$ & 2$\PZ$2$\PW$ & 422={\bf{211}}$\times$2 & 8\\
$u\bar u c\bar c c\bar s l\bar\nu$ & 2$\PZ$2$\PW$ & 422={\bf{211}}$\times$2 & 11 \\
$u\bar u s\bar s c\bar s l\bar\nu$ & 2$\PZ$2$\PW$ & 422={\bf{211}}$\times$2 & 11 \\
$u\bar u b\bar b c\bar s l\bar\nu$ & 2$\PZ$2$\PW$ & 233={\bf{211}}+22 & 15 \\
$d\bar d d\bar d c\bar s l\bar\nu$ & 2$\PZ$2$\PW$ & 422={\bf{211}}$\times$2 & 8\\
$d\bar d c\bar c c\bar s l\bar\nu$ & 2$\PZ$2$\PW$ & 422={\bf{211}}$\times$2 & 11 \\
$d\bar d s\bar s c\bar s l\bar\nu$ & 2$\PZ$2$\PW$ & 422={\bf{211}}$\times$2 & 11 \\
$d\bar d b\bar b c\bar s l\bar\nu$ & 2$\PZ$2$\PW$ & 233={\bf{211}}+22 & 15 \\
$c\bar c c\bar c c\bar s l\bar\nu$ & 2$\PZ$2$\PW$ & 1266={\bf{211}}$\times$6 & 5 \\
$c\bar c b\bar b c\bar s l\bar\nu$ & 2$\PZ$2$\PW$ &
466=({\bf{211}}+{\bf{22}})$\times$2 & 11 \\
$s\bar s s\bar s c\bar s l\bar\nu$ & 2$\PZ$2$\PW$ & 1266={\bf{211}}$\times$6 & 5 \\
$s\bar s b\bar b c\bar s l\bar\nu$ & 2$\PZ$2$\PW$ &
466=({\bf{211}}+{\bf{22}})$\times$2 & 11 \\
$b\bar b b\bar b c\bar s l\bar\nu$ & 2$\PZ$2$\PW$ &
610=({\bf{211}}+{\bf{22}}+{\bf{72}})$\times$2 & 8 \\
$u\bar u d\bar d c\bar s l\bar\nu$ & 2$\PZ$2$\PW$+4$\PW$ & 312={\bf{101}}+{\bf{211}} &
15\\
$c\bar c s\bar s c\bar s l\bar\nu$ & 2$\PZ$2$\PW$+4$\PW$ &
1046={\bf{101}}$\times$2+{\bf{211}}$\times$4 & 8 \\
%TOTAL &  &  & 141+20 \\
\hline
\end{tabular}\\[2pt]
\caption{Classification of 
$\Pp\Pp\rightarrow\Pq\Pq^\prime\rightarrow 4\Pq+\Pl\nu_{\Pl}$ processes.
The first column shows the group list, the second the process type,
the third the diagram number, and the last one the number of processes which
belong to the corresponding group. The numbers in boldface represent the 
independent sets of diagrams, as explained in the text.}
\end{table*}

\subsection{Process classification}
\label{sec:classes}

\pha is designed to compute SM processes with six fermions in the final state 
at the LHC. In this first version the code includes all \cal{O}($\alpha^6$) 
electroweak processes with four quarks, one lepton and one neutrino in the 
final state, $\Pp\Pp\rightarrow 4\Pq+\Pl\nu_{\Pl}$. More than one thousand 
processes belong to this class of final configurations, each one being
described by hundreds of diagrams. At first sight, the evaluation of such 
reactions appears rather daunting.
%More than one thousand processes belong to this class of final 
%configurations, and each process gets contributions from hundreds diagrams. 
%The evaluation of these reactions seems to be rather complicated at first 
%sight.
By making use of symmetries, the problem can be highly simplified. Taking into 
account one lepton type, charge conjugation and the symmetry between first and 
second family, 
the number of processes reduces to 161. A given reaction, its 
charge-conjugate, and the ones related by family exchange can be indeed 
described by the same matrix element; they differ by the Pdf's convolution. 
Moreover, all processes which share 
the same total particle content, with all eight partons taken to be outgoing, 
can be described by a single master amplitude.
%in the same way (they have different values but can be equally programmed).  
As a consequence, all thousand processes can be classified 
into 16 groups which are enumerated in Table 1. 
By selecting two initial quarks in each particle group, one obtains all
possible processes whose number is given in the last entry of the same table.
For example, from the particle set
$c\bar s d\bar u c\bar s l\bar\nu$ given in the first row of Table 1, where 
all fermions are by convention outgoing, one can derive the following 
processes:

\centerline{$\bar cs\rightarrow d\bar u c\bar s l\bar\nu\ \ \ \ \ \ 
\bar c\bar d\rightarrow\bar s\bar u c\bar s l\bar\nu\ \ \ \ \ \
\bar c u\rightarrow\bar s dc\bar s l\bar\nu\ \ \ \ \ \
\bar c\bar c\rightarrow \bar sd\bar u\bar s l\bar\nu$}
\centerline{$s\bar d\rightarrow c\bar u c\bar s l\bar\nu\ \ \ \ \ \
su\rightarrow cdc\bar s l\bar\nu\ \ \ \ \ \
ss\rightarrow cd\bar u cl\bar\nu\ \ \ \ \ \
\bar du\rightarrow c\bar s c\bar s l\bar\nu$}
\par\noindent
%After finding a key to organize processes and reduce them to a few groups 
%easy to be handled, the next step is working on diagrams. 
The calculation can be further simplified examining more closely the full set
of Feynman diagrams. The amplitudes of the aforementioned 16 groups are in 
fact not completely independent. One can show 
that they are combinations of just four basic sets of Feynman diagrams, 
composed of 101, 211, 22, 72 diagrams respectively (they are reported in table 
1 in boldface). This means that all thousand processes can be described using 
just a few building blocks. The immediate advantage is that any modification, 
like including new couplings or vertices, has to be performed only in a very 
restricted area of the program, and then it will be automatically 
communicated to all processes. The first two sets of 101 and 211 diagrams are 
related to the basic topologies the various processes can be classified in. In 
some processes, fermions can be paired only into charged currents (4W), giving 
rise to the first set of 101 Feynman diagrams. In some other process they 
can form two charged and two neutral currents (2Z2W), generating the 
second set of 211 diagrams. Mixed processes are described by a combination of
the two sets (2Z2W+4W). If a $\Pb$-pair is present, 2Z2W processes acquire
an additional set of 22 diagrams, describing Higgs contributions.
In case of two $\Pb$-pairs, 72 more diagrams are called in. This exhausts
the diagram classification for all processes with one neutrino in the final 
state, as we discuss in more detail in the next section. 
%The last two sets of 22 and 72 diagrams belong to the 
%2Z2W class, and correspond to Higgs contributions for processes with one or
%two {\it b}-pairs respectively.
%These three classes exhaust all six fermion processes with only one neutrino. 
\par\noindent
For every selected process \pha employs exact matrix elements, thus providing 
a complete description of signals and irreducible backgrounds. As an example, 
the reaction $c\bar c\rightarrow b\bar b c\bar s l\nu$ contains 
contributions coming from $\PW\PZ$-boson scattering, Higgs production with
subsequent decay to $\Pb\bar\Pb$ or $\PW\PW$, top pair 
and $\PW\PW\PZ$ production (where each of them can be considered either as a signal or as a background to the others). Since our approach 
is based on Feynman diagrams, it is possible to compute subsets of diagrams for a given amplitude. 
%This procedure is not in general gauge-invariant, and should be used with 
%extreme care expecially at the high LHC energies where strong gauge 
%cancellations can take place. Nonetheless, it might be useful in practice in 
%order to extract informations on the behavior of specific signals. 
In \phanosp, this possibility has been exploited for the Higgs diagrams, which 
can be switched off by the user. For practical details on how to select the 
different options we refer to \refse{sec:commoninput}. 
%Once the process has been chosen, the initialization subroutine looks at its 
%topology and automatically defines the kind of channels and corresponding 
%mappings to be used in the phase-space generation routine.

\subsection{Matrix elements}
\label{sec:matrixelements}

All amplitudes have been written with the help of the program {\tt PHACT} 
\cite{PHACT} ({\bf P}rogram for {\bf H}elicity {\bf A}mplitudes 
{\bf C}alculations with {\bf T}au matrices), which is based on the helicity 
formalism described in ref.\cite{helamp}.
This method is very powerful in coping with the complexity of
this kind of calculations. It is in fact based on a modular and diagrammatic 
approach. From the computational point of view, each Feynman diagram is not 
considered as a whole, but as a collection of several different pieces. 
One can thus independently compute parts of diagrams of increasing size and 
complexity, store them and assemble the various pieces only at the end. In 
this way, 
common subdiagrams are evaluated once, with a substantial efficiency gain.
In the following, we explain the method in a pictorial way, considering 
both 4W and 2Z2W processes. 
In computing any amplitude, one starts with the most elementary building 
blocks given by the subdiagrams corresponding to $\gamma$, $\PZ$, $\PW^\pm$ 
and Higgs boson decay into a pair of external fermions:

\bea
\begin{picture}(420,60)(0,0)
\thicklines
\SetOffset(30,30)
\Text(0,0)[]{$.$}
\Photon(0,0)(30,0){2}{3}
\Text(15,-14)[lb]{$\gamma$}
\ArrowLine(30,0)(45,15)
\Text(50,15)[l]{$p$}
\Line (30,0)(45,-15)
\Text(50,-15)[l]{$\ol p$}
\SetOffset(130,30)
\Photon(0,0)(30,0){2}{3}
\Text(15,-14)[lb]{$Z$}
\ArrowLine(30,0)(45,15)
\Text(50,15)[l]{$p$}
\Line (30,0)(45,-15)
\Text(50,-15)[l]{$\ol p$}
\SetOffset(230,30)
\Photon(0,0)(30,0){2}{3}
\Text(13,-14)[lb]{$W^\pm$}
\ArrowLine(30,0)(45,15)
\Text(50,15)[l]{$p$}
\Line (30,0)(45,-15)
\Text(50,-15)[l]{$\ol p'$}
\SetOffset(330,30)
\DashLine(0,0)(30,0){5}
\Text(15,-14)[lb]{$h$}
\ArrowLine(30,0)(45,15)
\Text(50,15)[l]{$p$}
\Line (30,0)(45,-15)
\Text(50,-15)[l]{$\ol p$}
\end {picture}
\eea

\noindent
Here and in the following, $p$ and $p^\prime$ indicate the isospin doublet 
components. By making use of these basic decays and of their insertions in a 
fermion line, one can then build the subdiagrams corresponding to a 
virtual $\gamma$, $\PZ$, $\PW^\pm$ or Higgs decaying into four outgoing 
fermions. For $\PW$-bosons we have:

\bea
\vcenter{\hbox{
\begin{picture}(80,60)(-5,-30)
\thicklines
\Photon(0,0)(30,0){2}{3}
\Text(10,-14)[lb]{$W$}
\Oval(45,0)(18,15)(0)
\Text(37,12)[lt]{$p$ $\ol p'$}
\Text(37,-12)[lb]{$q$ $\ol q$}
\end{picture}
}}
=
\vcenter{\hbox{
\begin{picture}(110,90)(-5,-45)
\thicklines
\Photon(0,0)(30,0){2}{3}
\Text(10,-14)[lb]{$W$}
\ArrowLine(30,0)(48,30)
\Text(40,40)[l]{$p$}
\Line (30,0)(48,-30)
\Text(40,-40)[l]{$\ol p'$}
\Photon(39,15)(69,15){2}{3}
\Text(46,0)[lb]{$\gamma ,Z$}
\ArrowLine(69,15)(84,30)
\Text(89,30)[l]{$q$}
\Line (69,15)(84,0)
\Text(89,0)[l]{$\ol q$}
\end{picture}
}}
+
\vcenter{\hbox{
\begin{picture}(110,90)(-5,-45)
\thicklines
\Photon(0,0)(30,0){2}{3}
\Text(10,-14)[lb]{$W$}
\ArrowLine(30,0)(48,30)
\Text(40,40)[l]{$p$}
\Line (30,0)(48,-30)
\Text(40,-40)[l]{$\ol p'$}
\Photon(39,-15)(69,-15){2}{3}
\Text(49,-30)[lb]{$\gamma ,Z$}
\ArrowLine(69,-15)(84,0)
\Text(89,0)[l]{$q$}
\Line (69,-15)(84,-30)
\Text(89,-30)[l]{$\ol q$}
\end{picture}
}}\nonumber \\
+
\vcenter{\hbox{
\begin{picture}(110,90)(-5,-45)
\thicklines
\Photon(0,0)(30,0){2}{3}
\Text(10,-14)[lb]{$W$}
\ArrowLine(30,0)(48,30)
\Text(40,40)[l]{$q$}
\Line (30,0)(48,-30)
\Text(40,-40)[l]{$\ol q$}
\Photon(39,15)(69,15){2}{3}
\Text(49,0)[lb]{$W$}
\ArrowLine(69,15)(84,30)
\Text(89,30)[l]{$p$}
\Line (69,15)(84,0)
\Text(89,0)[l]{$\ol p'$}
\end{picture}
}}
+
\vcenter{\hbox{
\begin{picture}(110,90)(-5,-45)
\thicklines
\Photon(0,0)(30,0){2}{3}
\Text(10,-14)[lb]{$W$}
\Photon(30,0)(45,25){2}{3}
\Text(26,12)[lb]{$W$}
\Photon(30,0)(45,-25){2}{3}
\Text(18,-24)[lb]{$\gamma ,Z$}
\ArrowLine(45,-25)(60,-10)
\Text(65,-10)[l]{$q$}
\Line (45,-25)(60,-40)
\Text(65,-40)[l]{$\ol q$}
\ArrowLine(45,25)(60,40)
\Text(65,40)[l]{$p$}
\Line (45,25)(60,10)
\Text(65,10)[l]{$\ol p'$}
\end{picture}
}}
+
\vcenter{\hbox{
\begin{picture}(110,90)(-5,-45)
\thicklines
\Photon(0,0)(30,0){2}{3}
\Text(10,-14)[lb]{$W$}
\Photon(30,0)(45,25){2}{3}
\Text(26,12)[lb]{$W$}
\DashLine(30,0)(45,-25){5}
\Text(27,-24)[lb]{$h$}
\ArrowLine(45,-25)(65,-10)
\Text(70,-10)[l]{$q$}
\Line (45,-25)(65,-40)
\Text(70,-40)[l]{$\ol q$}
\ArrowLine(45,25)(65,40)
\Text(70,40)[l]{$p$}
\Line (45,25)(65,10)
\Text(70,10)[l]{$\ol p'$}
\end{picture}
}}
\eea

\noindent
In the lower left-hand corner, the rightmost $\PW$-boson can be attached 
either on $\Pq$ or $\bar\Pq$, depending on its charge. According to the type 
of four-particle state (2W or 2Z), the subdiagrams corresponding to a virtual 
$\PZ$ or $\gamma$ decaying into four outgoing fermions are instead given by:

\bea
\vcenter{\hbox{
\begin{picture}(70,60)(-5,-30)
\Photon(0,0)(30,0){2}{3}
\Text(2,-16)[lb]{$\gamma (Z)$}
\Boxc(42,0)(24,30)
\Text(34,12)[lt]{$p$ $\ol p'$}
\Text(34,-12)[lb]{$q$ $\ol q'$}
\end{picture}
}}
=
\vcenter{\hbox{
\begin{picture}(110,90)(-5,-45)
\Photon(0,0)(30,0){2}{3}
\Text(2,-16)[lb]{$\gamma (Z)$}
\ArrowLine(30,0)(48,30)
\Text(40,40)[l]{$p$}
\Line (30,0)(48,-30)
\Text(40,-40)[l]{$\ol p'$}
\Photon(39,15)(69,15){2}{3}
\Text(49,0)[lb]{$W$}
\ArrowLine(69,15)(84,30)
\Text(89,30)[l]{$q$}
\Line (69,15)(84,0)
\Text(89,0)[l]{$\ol q'$}
\end{picture}
}}
+
\vcenter{\hbox{
\begin{picture}(110,90)(-5,-45)
\Photon(0,0)(30,0){2}{3}
\Text(2,-16)[lb]{$\gamma (Z)$}
\ArrowLine(30,0)(48,30)
\Text(40,40)[l]{$p$}
\Line (30,0)(48,-30)
\Text(40,-40)[l]{$\ol p'$}
\Photon(39,-15)(69,-15){2}{3}
\Text(54,-28)[lb]{$W$}
\ArrowLine(69,-15)(84,0)
\Text(89,0)[l]{$q$}
\Line (69,-15)(84,-30)
\Text(89,-30)[l]{$\ol q'$}
\end{picture}
}}+\hsk\hsk\hsk\nonumber \\
\hsk\hsk\hsk
\vcenter{\hbox{
\begin{picture}(110,90)(-5,-45)
\Photon(0,0)(30,0){2}{3}
\Text(2,-16)[lb]{$\gamma (Z)$}
\ArrowLine(30,0)(48,30)
\Text(40,40)[l]{$q$}
\Line (30,0)(48,-30)
\Text(40,-40)[l]{$\ol q'$}
\Photon(39,15)(69,15){2}{3}
\Text(49,0)[lb]{$W$}
\ArrowLine(69,15)(84,30)
\Text(89,30)[l]{$p$}
\Line (69,15)(84,0)
\Text(89,0)[l]{$\ol p'$}
\end{picture}
}}
+
\vcenter{\hbox{
\begin{picture}(110,90)(-5,-45)
\Photon(0,0)(30,0){2}{3}
\Text(2,-16)[lb]{$\gamma (Z)$}
\ArrowLine(30,0)(48,30)
\Text(40,40)[l]{$q$}
\Line (30,0)(48,-30)
\Text(40,-40)[l]{$\ol q'$}
\Photon(39,-15)(69,-15){2}{3}
\Text(54,-28)[lb]{$W$}
\ArrowLine(69,-15)(84,0)
\Text(89,0)[l]{$p$}
\Line (69,-15)(84,-30)
\Text(89,-30)[l]{$\ol p'$}
\end{picture}
}}
+
\vcenter{\hbox{
\begin{picture}(110,90)(-5,-45)
\Photon(0,0)(30,0){2}{3}
\Text(2,-16)[lb]{$\gamma (Z)$}
\Photon(30,0)(45,25){2}{3}
\Text(25,16)[lb]{$W$}
\Photon(30,0)(45,-25){2}{3}
\Text(25,-26)[lb]{$W$}
\ArrowLine(45,-25)(60,-10)
\Text(65,-10)[l]{$q$}
\Line (45,-25)(60,-40)
\Text(65,-40)[l]{$\ol q'$}
\ArrowLine(45,25)(60,40)
\Text(65,40)[l]{$p$}
\Line (45,25)(60,10)
\Text(65,10)[l]{$\ol p'$}
\end{picture}
}}
\eea

\noindent
and 

\bea
\vcenter{\hbox{
\begin{picture}(70,60)(-5,-30)
\Photon(0,0)(30,0){2}{3}
\Text(2,-16)[lb]{$\gamma (Z)$}
\Boxc(42,0)(24,30)
\Text(34,12)[lt]{$p$ $\ol p$}
\Text(34,-12)[lb]{$q$ $\ol q$}
\end{picture}
}}
=
\vcenter{\hbox{
\begin{picture}(110,90)(-5,-45)
\Photon(0,0)(30,0){2}{3}
\Text(2,-16)[lb]{$\gamma (Z)$}
\ArrowLine(30,0)(48,30)
\Text(40,40)[l]{$p$}
\Line (30,0)(48,-30)
\Text(40,-40)[l]{$\ol p$}
\Photon(39,15)(69,15){2}{3}
\Text(40,0)[lb]{$Z,\gamma ,h$}
\ArrowLine(69,15)(84,30)
\Text(89,30)[l]{$q$}
\Line (69,15)(84,0)
\Text(89,0)[l]{$\ol q$}
\end{picture}
}}
+
\vcenter{\hbox{
\begin{picture}(110,90)(-5,-45)
\Photon(0,0)(30,0){2}{3}
\Text(2,-16)[lb]{$\gamma (Z)$}
\ArrowLine(30,0)(48,30)
\Text(40,40)[l]{$p$}
\Line (30,0)(48,-30)
\Text(40,-40)[l]{$\ol p$}
\Photon(39,-15)(69,-15){2}{3}
\Text(46,-28)[lb]{$Z,\gamma ,h$}
\ArrowLine(69,-15)(84,0)
\Text(89,0)[l]{$q$}
\Line (69,-15)(84,-30)
\Text(89,-30)[l]{$\ol q$}
\end{picture}
}}+\hsk\hsk\hsk\nonumber \\
\hsk\hsk\hsk
\vcenter{\hbox{
\begin{picture}(110,90)(-5,-45)
\Photon(0,0)(30,0){2}{3}
\Text(2,-16)[lb]{$\gamma (Z)$}
\ArrowLine(30,0)(48,30)
\Text(40,40)[l]{$q$}
\Line (30,0)(48,-30)
\Text(40,-40)[l]{$\ol q$}
\Photon(39,15)(69,15){2}{3}
\Text(42,0)[lb]{$Z,\gamma ,h$}
\ArrowLine(69,15)(84,30)
\Text(89,30)[l]{$p$}
\Line (69,15)(84,0)
\Text(89,0)[l]{$\ol p$}
\end{picture}
}}+
\vcenter{\hbox{
\begin{picture}(110,90)(-5,-45)
\Photon(0,0)(30,0){2}{3}
\Text(2,-16)[lb]{$\gamma (Z)$}
\ArrowLine(30,0)(48,30)
\Text(40,40)[l]{$q$}
\Line (30,0)(48,-30)
\Text(40,-40)[l]{$\ol q$}
\Photon(39,-15)(69,-15){2}{3}
\Text(46,-28)[lb]{$Z,\gamma ,h$}
\ArrowLine(69,-15)(84,0)
\Text(89,0)[l]{$p$}
\Line (69,-15)(84,-30)
\Text(89,-30)[l]{$\ol p$}
\end{picture}
}}
+
\vcenter{\hbox{
\begin{picture}(110,90)(-5,-45)
\Photon(0,0)(30,0){2}{3}
\Text(8,-16)[lb]{$Z$}
\Photon(30,0)(45,25){2}{3}
\Text(24,16)[lb]{$Z$}
\DashLine(30,0)(45,-25){5}
\Text(24,-26)[lb]{$h$}
\ArrowLine(45,-25)(65,-10)
\Text(70,-10)[l]{$q$}
\Line (45,-25)(65,-40)
\Text(70,-40)[l]{$\ol q$}
\ArrowLine(45,25)(65,40)
\Text(70,40)[l]{$p$}
\Line (45,25)(65,10)
\Text(70,10)[l]{$\ol p$}
\end{picture}
}}
\eea

\noindent
%For mixed states, of the kind $p\bar pp^\prime\bar{p^\prime}$, both groups of 
%subdiagrams (2.3) and (2.4) contribute. 
Diagrams with a Higgs attached to a fermion line are computed only when 
$\Pb$-quarks are present. 
Finally, for the Higgs decay into four particles
%The remaining set of significative subdiagrams corresponds to the 
%Higgs decay into four particles. As before, 
we have two possible sets:

\bea
\vcenter{\hbox{
\begin{picture}(70,60)(-5,-30)
\DashLine(0,0)(30,0){5}
\Text(2,-16)[lb]{$h$}
\Boxc(42,0)(24,30)
\Text(34,12)[lt]{$p$ $\ol p'$}
\Text(34,-12)[lb]{$q$ $\ol q'$}
\end{picture}
}}
=
\vcenter{\hbox{
\begin{picture}(110,90)(-5,-45)
\DashLine(0,0)(30,0){5}
\Text(2,-16)[lb]{$h$}
\Photon(30,0)(45,25){2}{3}
\Text(22,16)[lb]{$W$}
\Photon(30,0)(45,-25){2}{3}
\Text(22,-28)[lb]{$W$}
\ArrowLine(45,-25)(60,-10)
\Text(65,-10)[l]{$q$}
\Line (45,-25)(60,-40)
\Text(65,-40)[l]{$\ol q'$}
\ArrowLine(45,25)(60,40)
\Text(65,40)[l]{$p$}
\Line (45,25)(60,10)
\Text(65,10)[l]{$\ol p'$}
\end{picture}
}}
\eea

\noindent
and

\bea
\vcenter{\hbox{
\begin{picture}(70,60)(5,-30)
\DashLine(0,0)(20,0){5}
\Text(2,-16)[lb]{$h$}
\Boxc(33,0)(24,30)
\Text(27,12)[lt]{$p$ $\ol p$}
\Text(27,-12)[lb]{$q$ $\ol q$}
\end{picture}
}}
=
\vcenter{\hbox{
\begin{picture}(110,90)(-5,-45)
\DashLine(0,0)(23,0){5}
\Text(2,-16)[lb]{$h$}
\ArrowLine(23,0)(43,30)
\Text(35,40)[l]{$p$}
\Line (23,0)(43,-30)
\Text(35,-40)[l]{$\ol p$}
\Photon(34,15)(65,15){2}{3}
\Text(36,0)[lb]{$Z,\gamma ,h$}
\ArrowLine(64,15)(79,30)
\Text(84,30)[l]{$q$}
\Line (64,15)(79,0)
\Text(84,0)[l]{$\ol q$}
\end{picture}
}}
+
\vcenter{\hbox{
\begin{picture}(110,90)(-5,-45)
\DashLine(0,0)(23,0){5}
\Text(2,-16)[lb]{$h$}
\ArrowLine(23,0)(43,30)
\Text(35,40)[l]{$p$}
\Line (23,0)(43,-30)
\Text(35,-40)[l]{$\ol p$}
\Photon(34,-15)(64,-15){2}{3}
\Text(41,-28)[lb]{$Z,\gamma ,h$}
\ArrowLine(64,-15)(79,0)
\Text(84,0)[l]{$q$}
\Line (64,-15)(79,-30)
\Text(84,-30)[l]{$\ol q$}
\end{picture}
}}+\hsk\hsk\hsk\nonumber \\
\hsk\hsk\hsk
\vcenter{\hbox{
\begin{picture}(110,90)(-5,-45)
\DashLine(0,0)(23,0){5}
\Text(3,-14)[lb]{$h$}
\ArrowLine(23,0)(43,30)
\Text(35,40)[l]{$q$}
\Line (23,0)(43,-30)
\Text(35,-40)[l]{$\ol q$}
\Photon(34,15)(64,15){2}{3}
\Text(35,0)[lb]{$Z,\gamma ,h$}
\ArrowLine(64,15)(79,30)
\Text(84,30)[l]{$p$}
\Line (64,15)(84,0)
\Text(84,0)[l]{$\ol p$}
\end{picture}
}}+
\vcenter{\hbox{
\begin{picture}(110,90)(-5,-45)
\DashLine(0,0)(30,0){5}
\Text(3,-14)[lb]{$h$}
\ArrowLine(30,0)(48,30)
\Text(40,40)[l]{$q$}
\Line (30,0)(48,-30)
\Text(40,-40)[l]{$\ol q$}
\Photon(39,-15)(69,-15){2}{3}
\Text(47,-28)[lb]{$Z,\gamma ,h$}
\ArrowLine(69,-15)(84,0)
\Text(89,0)[l]{$p$}
\Line (69,-15)(84,-30)
\Text(89,-30)[l]{$\ol p$}
\end{picture}
}}
+
\vcenter{\hbox{
\begin{picture}(110,90)(-5,-45)
\DashLine(0,0)(30,0){5}
\Text(3,-14)[lb]{$h$}
\Photon(30,0)(45,25){2}{3}
\Text(18,16)[lb]{$Z,h$}
\Photon(30,0)(45,-25){2}{3}
\Text(18,-26)[lb]{$Z,h$}
\ArrowLine(45,-25)(60,-10)
\Text(65,-10)[l]{$q$}
\Line (45,-25)(60,-40)
\Text(65,-40)[l]{$\ol q$}
\ArrowLine(45,25)(60,40)
\Text(65,40)[l]{$p$}
\Line (45,25)(60,10)
\Text(65,10)[l]{$\ol p$}
\end{picture}
}}
\eea

\noindent
depending on the specific four-particle configuration. Using these sets of 
$1\rightarrow 2$ and $1\rightarrow 4$ particle subdiagrams as building blocks, 
the 4W-type amplitude assumes the extremely concise structure given in Fig. 1.
The full matrix element can be expressed just in terms of four main 
topologies. The second one drawn in the figure is described by two diagrams, 
as $\PW^+\PW^-$ pairs can be formed in two different ways. The third topology 
represents eight diagrams, as all four fermion pairs can emit three $\PW$'s 
and for each given fermion line one can 
\unitlength 1cm
\begin{picture}(15.,8.5)
\put(-2.,-10.){\includegraphics{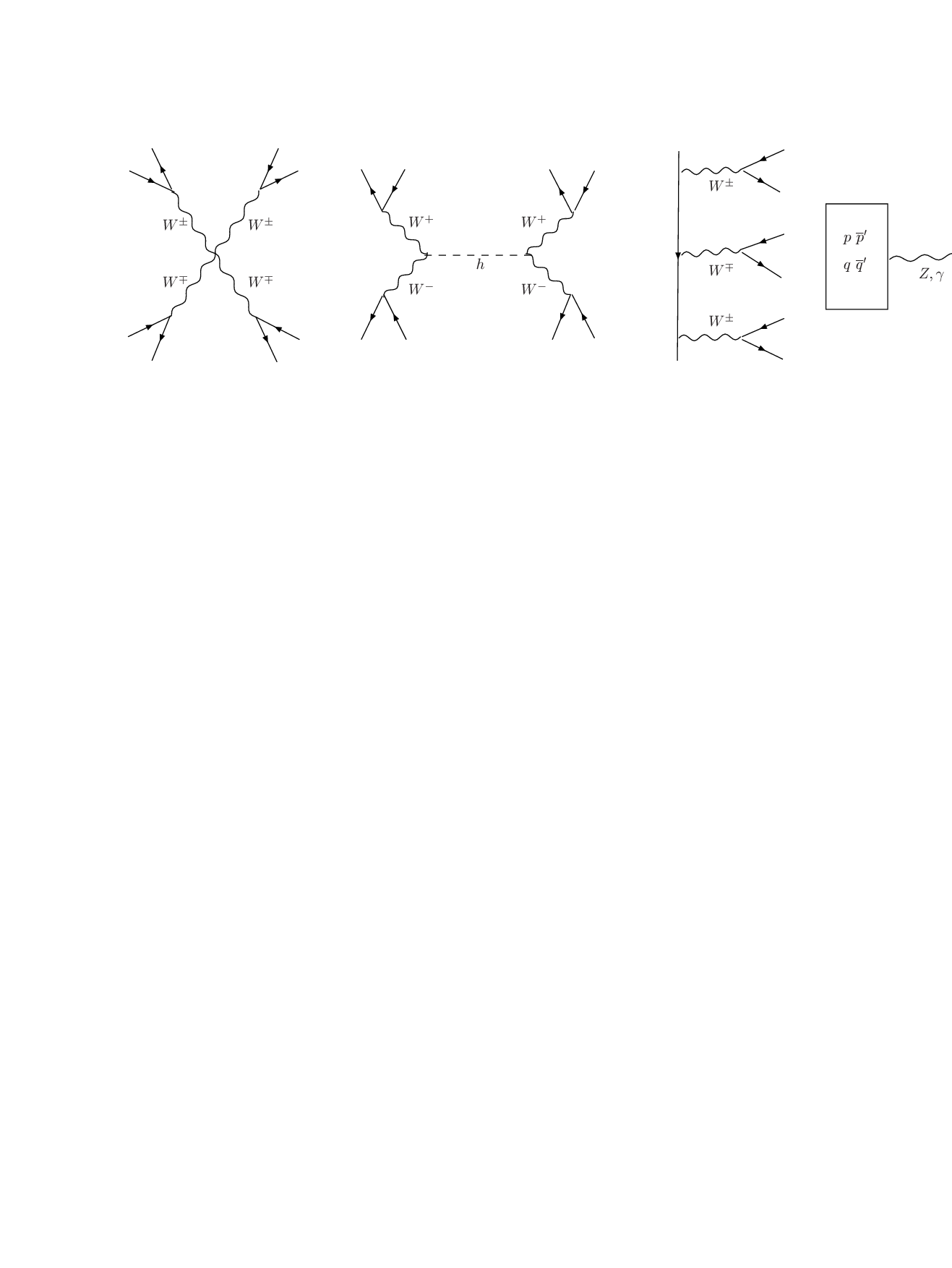}}
\end{picture}
\vskip -4.cm
\centerline{Figure 1: Diagrams for 4W-type processes of the kind 
$c\bar sd\bar uc\bar sl\bar\nu$}

\vskip 1.cm
\noindent
exchange the two like-sign vector bosons. 
Finally, 
the last graph includes ninety diagrams. In case of $\PZ$-boson exchange we have 
in fact five subdiagrams for each side, already summed up as shown in 
Eq.(2.3), and two different ways to form a $\PW^+\PW^-$ pair. 
In presence of one 
neutrino in the final state, which is the case we are addressing in this 
first version, the number of diagrams with $\gamma$ exchange gets reduced to 
forty. For 4W-type processes, we therefore end up with 101 basic diagrams, as 
reported in the first row of Table 1. 
Analogously, 2Z2W-type processes have the simple structure outlined in Fig. 2.
Considering Eqs.(2.2)-(2.6), one can easily see that these processes have 211
diagrams if there is no $\Pb$-quark. In presence of a 
$\Pb\bar\Pb$ pair, there are 22 additional diagrams which constitute a 
further independent set. Finally, one more separate set given by 72 
diagrams contributes to channels with four $\Pb$-quarks. Mixed processes need
both $4\PW$ and $2\PZ2\PW$ contributions. These two sets of diagrams
describe (unrealistic) processes where all fermions are different. They 
constitute the essential kernel, from which all other related diagrams can be 
derived. Additional diagrams accounting for identical particles are in fact 
simply obtained by fermion exchange. This explains the numbers reported in the 
third column of Table 1. These numbers are quoted only for reference, as we do 
not compute every single diagram but only the few topologies of Figs. 1,2.
\par\noindent
The helicity amplitude formalism is appropriate both for massless and massive 
fermions. At the present stage, fermion masses are taken into account for 
{\it bottom} and {\it top} fermion lines. This strategy provides an excellent 
approximation to the full result in all cases which do not exhibit collinear 
or mass singularities. In this first version, we aim to cover all possible
processes $\Pp\Pp\rightarrow 4\Pq+\Pl\nu_{\Pl}$ with hard and well separated 
fermions in the final state. Full 
massive amplitudes would be however just a straightforward extension of the 
code, with the only drawback of slowering the program. The number of helicity 
states increases and new terms appear in the diagram evaluation. However the 
logic of constructing progressively the building blocks stays unaltered. 
\pha is 
structured in such a way that makes it easy to accomodate possible future 
developments.   
\par\noindent
\pha matrix elements, squared and summed over polarizations, have been 
extensively compared with {\tt Madgraph}\cite{Madgraph} amplitudes. A very good numerical agreement has been found. To conclude this section, let us briefly 
comment on the inclusion of weak boson finite-width effects. As well known, 
this requires a careful treatment. It is in fact closely related to 
\unitlength 1cm
\begin{picture}(15.,8.5)
\put(-2.,-10.){\includegraphics{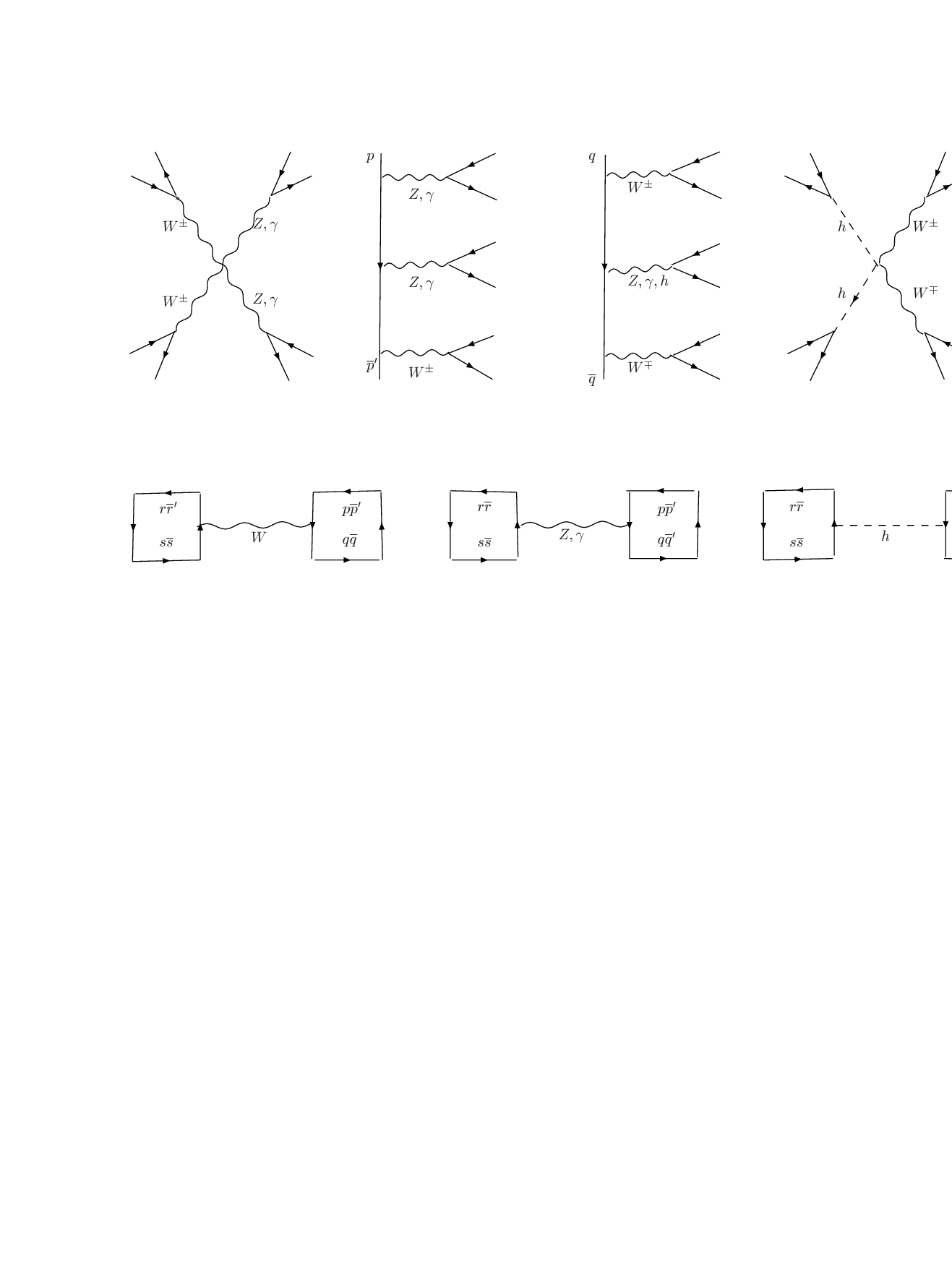}}
\end{picture}
\vskip -.6cm
\centerline{Figure 2: Diagrams for 2Z2W-type processes of the kind 
$b\bar bb\bar bc\bar sl\bar\nu$.}
\vskip 1.cm
\noindent
the gauge invariance of the theory, and even 
tiny violations of Ward identities can lead to totally wrong results in many 
cases. There are several schemes in the literature for the introduction of the 
decay width in the propagators. The most appealing approach is the 
Fermion-Loop scheme \cite{fl}, which preserves gauge invariance. It however 
requires the computation of a considerable number of additional terms in the 
amplitude. An alternative simpler option is the fixed-width scheme (FW). 
In the unitary gauge we work in, it consists in replacing $M^2$ with 
$M^2-iM\Gamma$ both in the denominator and in the $p^\mu p^\nu$ term of the 
vector boson propagator. This scheme preserves U(1) gauge invariance at the 
price of introducing unphysical widths for space-like vector bosons.
In \pha we have chosen to implement this latter scheme. 

%\subsection{Phase space sampling}
\subsection{Iterative-adaptive multichannel}
\label{sec:phasespace}

In this section a new integration method is described. It employs an iterative 
and adaptive multichannel technique. The ability to adapt is the overriding 
consideration for multidimensional integrals of discontinuous and sharply 
peaked functions. 
\par\noindent
Computing a six-fermion process at hadron colliders requires an integration 
over a 16-dimensional space. The generic process can be written as
\be
h_1 + h_2 \to f_3 + f_4 + f_5 + f_6 + f_7 + f_8 + X
\ee 
where $h_1$ and $h_2$ denote the incoming protons, $f_i$ the outgoing 
fermions, and $X$ the remnants of the protons. In the parton model the 
corresponding cross sections are obtained from the following convolution 
\ba\label{eq:convolution}
\si^{h_1 h_2}(P_1,P_2,p_f) = \sum_{i,j}\int_0^1\rd x_1 \int_0^1\rd x_2\,
F_{i,h_1}(x_1,Q^2)F_{j,h_2}(x_2,Q^2) \,
\int_{\Phi_{6f}}\rd\hat\si^{ij}(x_1P_1,x_2P_2,p_f) 
\nonumber
\ea
\ba
\int_{\Phi_{6f}}\rd\hat\si^{ij}&=&
\frac{1}{2\hat{s}}\int_{\Phi_{6f}}\prod^8_{i=3}{\rd^3{\bf p}_i\over{2p_i^0}}
\delta^{(4)}\left (x_1P_1+x_2P_2-\sum^8_{j=3}p_j\right )
{|M(x_1P_1,x_2P_2,p_f)|^2\over{(2\pi )^{14}}}
\ea
\vskip 0.3cm
\noindent
where $p_f$ summarizes the final-state momenta, $F_{i,h_1}$ and
$F_{j,h_2}$ are the distribution functions of partons $i$ and $j$ in the 
incoming protons $h_1$ and $h_2$ with momenta $P_1$ and $P_2$, respectively, 
$Q$ is the factorization scale, and $\hat\si^{ij}$ represent the cross 
sections for the partonic processes averaged over colours and spins of the 
partons. The sum $\sum_{i,j}$ runs over all possible quarks 
$\Pu ,\Pd ,\Pc,\Ps,\Pb$. 
Finally, the symbol $\Phi_{6f} $ denotes the six-particle phase space and 
$\hat{s}=(x_1P_1+x_2P_2)^2$ the center of mass (CM) energy squared in the 
partonic system.
\par\noindent
Integrating numerically eq.\refeq{eq:convolution} is rather complicated. An 
individual process can contain hundreds of diagrams. The resonant peaking 
structure of the amplitude is therefore generally very rich. As a consequence 
the 16-dimensional space has non-trivial kinematical regions corresponding
to the enhancements of the matrix elements. In a fully extrapolated setup,
these peaks are simultaneously present. Of course, the requirement of suitable 
cuts can enhance some resonances while supressing others. Our aim is to 
have maximal coverage of phase space so as to fully exploit the LHC 
potential in measurements and searches. 
%We thus want to work with the loosest set of cuts. 
Given the complexity of the final state, it is necessary to 
develop a reliable and efficient phase-space sampling algorithm.
%in order to reach even the most delicate kinematical regions where problems 
%of numerical instabilities could easily lead to inconsistent results, if not 
%properly treated. 
\par\noindent
Two are the most advanced and commonly adopted integration techniques:
the {\it multichannel} method \cite{multichannel} and the {\it adaptive} 
approach \`a la {\tt VEGAS} \cite{vegas}. 
%They both rely on the basic concept 
%of {\it importance sampling} for reducing the integration error. In this 
%approach, (pseudo)random variables $0\le x_i\le 1$ are generated with a 
%non-uniform distribution in order to have a higher density of Monte Carlo 
%points in those regions where the integrand is largest in magnitude. However, 
The two strategies are completely different. In the {\it multichannel} 
approach, mappings into phase-space variables are chosen in such a way that 
the corresponding Jacobians cancel the peaks of the differential cross section. These mappings are not in general unique. One normally needs several different 
phase-space parametrizations, called channels, one for each possible peaking 
structure of the amplitude. In principle every single diagram can have a 
different resonant pattern described by a different set of variables.
In addition, some variables corresponding to an individual diagram can 
resonate or not. This gives rise to a huge combinatorics which requires a 
correspondingly large number of channels. 
%The reason is twofold.
%A phase space parametrization is uniquely defined by the choice of the set of 
%variables, which describe a certain resonant structure. However, matrix 
%elements can show peaks in different regions of phase space that are better
%described in terms of different sets of variables. In addition, for any given 
%set, one has to keep in mind the intrinsic combinatorics of the 
%amplitude peaking structure. Depending on the phase space region, one or more 
%variables could either resonate or not, giving rise to a multiplicity of 
%possible mappings which increases with the number of particles. 
%Number and type of the various channels must be fixed a priori, 
Number and type of these mappings must be fixed a priori, before starting the 
integration. The {\it multichannel} method thus requires a guess on the 
behavior of 
the integrand function. It indeed relies on the expectation that the selected
set of channels, properly weighted \cite{weights}, is able to describe 
reasonably well the amplitude. As no adaptivity is provided (a part from 
the freedom to vary the relative weight of the different channels), neglecting 
even one channel might worsen considerably the convergence of the integral. 
It is thus clear that, as the number of particles increases, the use of 
this technique becomes rather cumbersome and time consuming.
%For this reason, the question of its efficiency in the 
%event generation is still matter of discussion. Up to now, there are very few 
%applications of the {\it multichannel} method for phenomenological studies of 
%processes with six or more partons in the final state.
\par\noindent
The criteria of {\it adaptive} integration as performed by {\tt VEGAS} are 
rather different. This approach bases its strenght on the ability to deal 
automatically with 
totally unknown integrands. By employing an iterative method, it acquires 
knowledge about the integrand during integration, and adapts consequently 
its phase-space grid in order to concentrate the function
evaluations in those regions where it peaks more. In this case, the capability 
of adapting well to the function while integrating depends on two 
factors: the choice of phase-space variables and the binning refinement.
{\tt VEGAS} divides the N-dimensional space in hypercubes, and scans the 
integrand along the axes. For a good convergence of the integral, it thus 
requires amplitude peaks to be aligned with the axes themselves. The problem 
can be easily solved if one set of phase-space variables is sufficient 
to describe the full amplitude peaking structure. In this case, the alignement 
can always be obtained by an appropriate variable trasformation. The method 
becomes inefficient when it is impossible to align all enhancements with a 
single trasformation.
%more trasformations are needed to take care of different peaking topologies. 
This is the main weakness of the adaptive algorithm. In addition, if the 
binning is too coarse, some narrow peaks can excape detection, even if along the axes, with 
consequent instability or underestimation of the integral.
The two approaches have clearly complementary advantages and disadvantages. 
\par\noindent
We have devised a new integration method, called 
{\it iterative-adaptive multichannel}, which merges the multichannel strategy 
with the iterative-adaptive approach. An algorithm based on a similar 
philosophy has been proposed in \cite{vamp}.
%It improves both methods, while preserving their best features. 
%Very briefly summarizing, that consists in 
%a multichannel unbounded to have necessarily all possible channels specified, 
%and where the unpredictable behaviour of the integrand is left to adaptivity. 
Our integration method makes use of the {\tt VEGAS} routine.   
It is characterized by two main features, named {\it multi-mapping} and 
{\it {\tt VEGAS}-multichannel}, which we are going to describe in the next
two sections. The first one aims at reducing the number of separate channels 
one has to consider in the {\it multichannel}. 
The latter provides the necessary adaptivity.
 
\subsubsection{Multi-mapping}
\label{sec:mapping}

%The function of the {\it multi-mapping} is twofold. It allows to enclose more 
%phase-space parametrizations in the same channel, reducing the effective
%number of independent channels to be integrated on. And, it improves
%considerably the {\tt VEGAS} adaptivity. This helps in concentrating 
%phase-space grids in those regions where the integrand peaks more.
In this section, we describe how the integrand peaking structure gets smoothed 
through the employment of proper random number mappings into phase-space 
variables. We suppose to have a unique phase-space parametrization defined by 
a certain set of variables. A typical example of amplitude enhancement is 
given by a bosonic resonance decaying into two particles.   
%In this section, we discuss how to enhance {\tt VEGAS} adaptivity through the 
%employment of proper random number mappings into phase-space variables.
%This helps in concentrating phase-space grids in those regions where the 
%integrand peaks more. To this end, we suppose to have a unique phase space
%defined by a certain set of variables, and we describe how the integrand 
%peaking structure get smoothed.
%smooth resonant peaking structures with proper phase space parametrizations. 
%In \pha two types of mapping are essentially used: Breit-Wigner resonant
%mapping (s-channel) and small angle mapping (t-channel).
%; we briefly sketch them in the following. 
%Unless dominated by gluons, six-fermion final states contain at least a pair 
%of particles which could come from the decay of a $W$, $Z$ or Higgs boson. 
Let us take for instance the case of a fermion pair $f_i\bar{f_j}$. A natural 
variable is then the invariant mass $m_{ij}=\sqrt{(p_i+p_j)^2}$. Whenever 
$m_{ij}$ is close to the mass of a $\PW$, $\PZ$ or Higgs boson, the 
corresponding amplitude squared shows a Breit-Wigner resonance. 
%Assuming all remaining variables to be flat, 
The total integral can be represented as
\be
I = \int d\Phi \frac{f^\prime (\Phi )}{\left(m_{ij}^2-M_B^2\right)^2+{M_B^2\Gamma_B^2}}~~~~~~~~~~B=W,Z,H
\label{integral}
\ee
where $\Phi$ is the full set of phase-space variables, including $m_{ij}$, and 
$f^\prime (\Phi )$ a smooth function of $m_{ij}$.
It is therefore convenient to perform a variable trasformation and use, instead of the invariant mass $m_{ij}$, an integration variable proportional to 
\be\label{tang}
x\propto\Parctan\left (\frac{m_{ij}^2-M_B^2}{M_B\Gamma_B}\right )
\label{var}
\ee
%(each variable $x$ is always transformed to the interval $0 \div 1$). 
In this way, the integral in eq.\refeq{integral} can be written as
\be
I = \int d\Phi ~g(\Phi ) ~f^\prime (\Phi ) = \int dx~f^{\prime\prime} 
(\Phi (x))
\ee
$g(\Phi )$ being the non-uniform probability density according to which 
phase-space variables are distributed. The function $f^{\prime\prime}$ is
given by $f^{\prime\prime}=f^\prime (\Phi )/(2m_{ij}M_B\Gamma_B)$. The 
Jacobian of the $\Phi\rightarrow x$ trasformation cancels the Breit-Wigner 
peak. We refer to (\ref{tang}) as resonant mapping. The example we just 
discussed is very 
simple, and often inadequate to deal with the actual complexity of matrix 
elements. One can have in fact more peaks appearing on the same variable, and 
long not-resonant tails which extend far away from the peaks. This latter case 
is more and more severe as the collider energy increases. To solve this 
problem, we have introduced {\it multi-mapping}. Let us consider the case of a 
neutral fermion pair $f_i\bar{f_j}$ which could originate from the decay of a 
$\PZ$ or a Higgs boson. For this particle configuration, a double mapping of 
the 
type (\ref{tang}) is performed simultaneously on $m_{ij}$. In order to cover 
all not-resonant regions, a uniform mapping (flat) is employed in the 
remaining integration range. In this particular case and assuming $M_H>M_Z$, 
we end up with five integration domains and three corresponding mappings: 
flat, $\PZ$-resonant, flat, Higgs-resonant, flat. This is what we call 
{\it multi-mapping} on a given variable. 
%Of course, the existance of all possible integration sub-intervals depends on 
%available energy and applied cuts. The phase space generation is optimized 
%for each individual hard process, including the selected cuts directly on the 
%integration limits of our variable $m_{ij}$.   
The advantage of using a {\it multi-mapping} is that the same channel can 
enclose several phase-space parametrizations, with a substantial gain in 
efficiency and CPU time. The number of separate channels decreases 
considerably.
Just to give an idea of the mapping combinatorics, let us consider the decay 
of a neutral boson into four fermions 
$B\rightarrow f\bar ff^\prime\bar{f^\prime}$. Among all possible integration 
variables, we can choose the three invariant masses $m_B$, $m_{f\bar f}$ and 
$m_{f^\prime\bar{f^\prime}}$. 
%(another choice could be the set $m_B$, $m_{f\bar ff^\prime}$ and 
%$m_{f\bar f}$). 
With three mappings per variable, as we said before, this generates 27 
mappings. In a standard multichannel, 27 distinct channels would be required.
In our approach, a single channel can cover all different kinds of 
triply, doubly, singly and not-resonant topologies, relying on integration 
adaptivity. 
\par\noindent
The previous example of resonant multi-mapping does not exhaust all possible  
amplitude peaking structure. For instance, one can have narrow peaks also in 
t-channel propagators, when one of the outgoing particle is emitted in the 
forward/backward direction with respect to the beam. The mapping for these 
small angle regions is inspired to the method of Ref. \cite{smallangle}. 
{\it Multi-mapping} applies here as well. The range of the phase-space 
variable, which is the denominator of the t-channel propagator, can be divided 
into two regions. One corresponds to small scattering angles, and is mapped to 
a power-like behaviour. The other one is related to a possibly large
not-resonant range. Both types of multi-mapping,
acting on different phase-space variables, can be combined within the same
channel to describe the most relevant part of the amplitude peaking structure.
Integration adaptivity takes care of the residual discrepancy between our 
parametrization and the actual behaviour of the amplitude.  

\subsubsection{{\tt VEGAS} multichannel}
\label{sec:multichannel}

While {\it multi-mapping} is extremely useful to improve the convergence of 
{\tt VEGAS} integration within a single phase-space parametrization, in
general several such parametrizations are needed. In this case, one has to
%{\it Multi-mapping} alone would be sufficient to improve the convergence of 
%the {\tt VEGAS} integral, in presence of just one set of variables 
%representing the full amplitude structure. If more sets are needed, one has 
%to introduce $N$ different phase-space parametrizations (or channels in 
%standard multichannel language) each one with its proper {\it multi-mapping}. 
introduce $N$ different channels (in standard multichannel language) with 
their proper {\it multi-mapping}. Each channel defines a non-uniform 
probability density $g_i(\Phi )$, which describes a specific class of 
amplitude peaks.
%according to which phase-space variables are distributed. 
%This should be chosen in such a way to best describe a specific class of 
%amplitude peaks. 
If we had just a single channel, as in the previous section, denoting with 
$f(\Phi )$ the function to be integrated, we would write
\bq
I = \int d\Phi f(\Phi ) = \int d\Phi ~g(\Phi ){f(\Phi )\over{g(\Phi )}}= 
\int d\Phi ~g(\Phi )f^\prime (\Phi ) = \int dx~f^\prime (\Phi )
\label{multi}
\eq
\par\noindent
where $f^\prime (\Phi )$ represents the smooth part of the integrand, once the 
peaking structure has been canceled. In presence of $N$ channels, the sum of 
the probability densities, properly weighted, should give the best description 
of the matrix element squared. Generalizing eq.\refeq{multi} to a number $N$ 
of channels, one can then write
\bq
I = \int d\Phi f(\Phi ) = \sum_{i=1}^N\alpha_i\int {d\Phi~g_i(\Phi)f(\Phi )
\over{\sum_{i=1}^N\alpha_i~g_i(\Phi )}} = \sum_{i=1}^N\alpha_i\int 
dx_i~f^{\prime}(G^{-1}_i(x_i)) = \sum_{i=1}^N\alpha_iI_i
\label{multigen}
\eq
\par\noindent
$\alpha_i$ being the so called {\it weight} of the {\it i}-th channel 
($x_i=G_i(\Phi )$), and $f^{\prime}(\Phi )$ the smoothed integrand. The 
$\alpha_i$ quantify the relevance of the different peaking structures of the 
amplitude. They must be chosen reasonably well in order to fit the integrand, 
i.e. to obtain a well behaved function $f^{\prime}(\Phi )$.
Owing to the very poor knowledge of the integrand, it is rather difficult to 
guess these values a priori. Usually, they are computed and optimized during 
the integration run. The algorithm described in eq.\refeq{multigen} is nothing
else than the standard multichannel.
In this method, the integral is computed in a single run, picking up the 
various channels with probability given by the corresponding $\alpha_i$ weight.
In the {\it iterative-adaptive multichannel}, the integral in 
eq.\refeq{multigen} splits in $N$ distinct contributions. 
The presence of identical final-state particles increases the possible list of
resonant structures. In order to keep the number of separate integration runs 
manageable, we include all jacobians generated by particle exchange in the
denominator of eq.\refeq{multigen}, while exploiting the freedom to relabel
the momenta to regroup all integration runs related by particle exchange to a 
single one. Owing to the 
{\it multi-mapping} technique previously described, and to the adaptivity of 
the integration algorithm, a maximum of seven channels is required to 
calculate all processes in Table 1. The criteria to automatically define 
number and type of channels needed for a given process are the following. 
We identify phase-space variables in which enhancements can appear due to 
boson and top propagators. We then consider the different sets of 
variables in which the maximum number of such propagators can be 
simultaneously present. These will determine our channels.
%We consider the maximum 
%number of boson and top propagator resonances which can simultaneously appear. %Fermion resonances are taken into account only for top quark. 
Multi-mapping and adaptivity will take care of all related partially-resonant 
or not-resonant configurations, as explained in \refse{sec:mapping}.
Each channel in eq.\refeq{multigen} is integrated separately with {\tt VEGAS}.
%In order to speed up the improvement of the grid, we perform a warm-up run, 
%called thermalization, with a relatively small number of points. These 
%preliminary fast evaluations are used only to adjust the grid, and give no 
%contribution to the final integral. 
In the {\it iterative-adaptive multichannel} method, a thermalization stage 
with a relatively small number of points is employed to determine the 
relative weights $\alpha_i$ of the various channels as follows. 
%We follow this procedure. 
In every thermalizing iteration, all channels are independently integrated 
for some set of $\alpha_i$. At the end of each iteration, a new set of 
phase-space grids (one for each channel), and an improved set of $\alpha_i$ 
are computed. The criteria for weight optimization we adopt is 
\be
\alpha_i={I_i\over{\sum_{i=1}^N I_i}}
\ee
\noindent
where $I_i$ is the $i$-channel integral. The new sets of $\alpha_i$ and grids 
are then used in the next iteration. The procedure is repeated until a good 
stability of the $\alpha_i$ is reached. In the standard multichannel method, 
the 
final result depends sensitively on the accuracy obtained for the $\alpha_i$ 
values. In our approach, owing again to integration adaptivity, only a rough 
estimate of the relative weights of the individual channels is sufficient for 
an accurate integration. Having established the relative weights and having 
obtained the initial grids, one can start the actual integration run, where 
the $N$ channels are evaluated in sequence. The iterative-adaptive algorithm 
is applied at this stage as well, and new grids are generated after each step. 
The last iteration produces $N$ final grids, which contain full information on 
the integrand function. The grids are stored in files, called in the 
following {\it grid-files}, and used whenever needed in the so called 
{\it one-shot} event generation, which we are going to describe in the next 
section. 

\subsection{Unweighted event generation}
\label{sec:eventgeneration}

Once phase-space grids are ready, the generation of unweighted events 
can start. 
This procedure, called {\it one-shot}, represents one of the main features of 
\phanosp. Inspired to the method used in ref.\cite{wphact}, it allows the user 
to generate 
unweighted events not on a process by process basis, but for all possible 
processes (or any selected subset of them) in just a single run. The result is 
a complete event sample, where all included final states appear in the right 
relative proportion. The algorithm is based on the {\it hit-or-miss} method. 
Thus, it needs to know the maximum value of the integrand functions of all
channels and processes.
When running in {\it one-shot} mode, all necessary informations about processes
are read from the {\it grid-files}, where they have been recorded during the 
grid preparation. In addition to the phase-space grid, these files contain 
also process and channel labels, the corresponding $\alpha_i$ weights, and the 
maximum value of the integrand function. Relying on these inputs, the code 
computes  
the probability according to which every single channel is picked up during 
the unweighted generation. Events will be generated using a modified version of {\tt VEGAS}, which chooses at random a cell of the phase-space grid read 
from the {\it grid-files}. A good determination of the grids traslates into
high efficiency. The procedure is repeated until 
the required number of unweighted events is produced.
\par\noindent
Every generated event may be either directly passed to {\tt PYTHIA} 
\cite{pythia}, for showering and hadronization, or can be stored into files 
for further processing. In this way, 
one has a complete and accurate tool for realistic experimental simulations. 
This step is performed according to the \LHP ~\cite{leshouches}, a set of 
common blocks for passing event configurations from parton level generators to 
parton shower and hadronization packages.

\section{Running modes}
\label{sec:modes}

\noindent
In this section we discuss how to run the code. We just give a guideline, 
useful to understand running mechanism and possible options. For a more 
detailed description we refer to Appendix \ref{sec:input}. The program 
has two modes of operation: {\it single-process} and {\it one-shot}, which 
%They represent two distinct stages of the same computation, and 
are selected by the input values {\tt ionesh=0,1} respectively. In the former 
mode ({\tt ionesh=0}), the code evaluates all $N$ processes one wants to 
generate in $N$ separate runs, and prepares the {\it grid-files}
%In each individual run, the program calculates, for every channel of each 
%process, the integrand maximum and phase-space grid 
to be used in the {\it one-shot} generation. In this latter mode 
({\tt ionesh=1}), the code generates instead 
unweighted events for all processes (or any subset of them specified by the 
user) in the same run.
%producing a complete sample where each final state appears in the correct 
%relative proportion. 
The two modes correspond to two 
distinct branches of the program. They thus need two different input sets.
Both sets are included in the same {\it input-file}. The first part of this
file is common to both modes. The rest depends on the 
selected mode. A practical feature of the input routine is that variables, 
which do not need to be specified in the chosen running mode, can be left in 
the {\it input-file} without harm. They are simply ignored.
The {\it input-file} must always be called {\tt r.in}. A sample {\tt r.in} is 
supplied with the program package. A detailed description, explaining meaning 
and possible values of input variables, is given in Appendix 
\ref{sec:input}.
In this section, we just discuss the computational strategy, and the main 
options which are available.

\subsection{Single-process}
\label{sec:singleprocess}

The use of this mode is twofold. The easiest option is computing the cross 
section of a specific process. This might be useful for some test or dedicated 
analysis. The alternative choice is to employ the {\it single-process} mode 
as necessary pre-run for the {\it one-shot} generation. In this case, the 
main purpose is the production of phase-space grids. This implies an extensive 
use of {\tt ionesh=0}, devoted to compute in separate 
runs all processes one intends to consider in the {\it one-shot} generation. 
A Perl script ({\tt setupdirSGE.pl}) for creating a tree-structure with
subdirectories and {\it input-files}, one for each process, is provided
in the program package.
For every single evaluation, the user must specify the desired reaction. The 
variable to fill is called {\tt iproc}, and uses the standard Monte Carlo 
particle numbering scheme, as described in Appendix \ref{sec:input}. 
Once the process has been selected, a first routine initializes parameters and
variables. It defines number and kind of channels appearing in the integration,
according to the algorithm discussed in \refse{sec:multichannel}. These 
informations, along with the corresponding phase-space parameters, are then 
passed to the phase-space generation routines, and lately to the integration 
algorithm. The integration routine is based on {\tt VEGAS}; it thus needs  
{\tt VEGAS} parameters to be defined. The user must specify integration
accuracy, number of iterations and Monte Carlo points per iteration,
both for thermalization and actual integration. As explained in 
\refse{sec:multichannel}, the program has an initial warm-up stage, followed by the actual process evaluation. For every single process, in thermalization the 
code determines the relative weight of each channel appearing in the 
multichannel integration, and produces a first instance of phase-space grids 
(one per channel). These grids are then used as a starting point for the second step, which consists of $M$ separate integrations, where $M$ is the number of 
channels. Each integration typically proceeds through several iterations. At 
the end of each iteration, the phase-space grid gets refined in an effort to 
decrease the overall variance. After the last iteration, the optimal grid is 
recorded in the {\it grid-file} named {\tt PHAVEGAS0i.DAT}, where {\tt i} 
represents the corresponding channel index.
%to be used later in the {\it one-shot} event generation. 
In the
same file, are also stored the maximum of the integrand function $w0$ produced 
in the next--to--last iteration, and the maximum $w1$ produced in the last 
iteration. 
%Phase-space grids and maxima are then used in {\it one-shot} mode, which 
%employes the {\it hit-or-miss} method for the unweighted event generation. 
\par\noindent
Before concluding this section, let us discuss the \pha options for imposing 
kinematical cuts. The {\it input-file} ({\tt r.in}) provides a predetermined 
set of kinematical cuts. Basically, two different types of cuts have been 
predisposed. A first set allows to approximately simulate detector acceptance 
and separation requirements. The second one allows to require two 
forward-backward jets, and two jets and one charged lepton in the central 
region. This signature helps to suppress QCD background, enhancing Higgs and 
vector boson scattering signals. 
The complete list of predetermined cuts, their meaning and logic are 
given in Appendix \ref{sec:input}. In addition, it is also possible to include 
extra user-specified cuts via a routine called {\tt IUSERFUNC}, an example of 
which is provided in the program package. This part of the input is common to 
both running modes, and must be always kept unchanged when passing from 
{\tt ionesh=0} to {\tt ionesh=1}. It constitutes in fact the setup under which 
phase-space grids are produced. In order to give the possibility of imposing 
other cuts at generation level, the {\it input-file} has also a cut section 
specific of the {\it one-shot} mode, which we describe in the next section. 
%A further possibility is imposing cuts at generation level,
%as explained in the next section.

\subsection{One-shot}

Once phase-space grids are ready, the {\it one-shot} mode allows the user to 
generate unweighted events for all processes simultaneously. The outcome is a 
complete event sample, able to simulate the full six-fermion production. 
%The generated sample will be in fact representative of all reactions, with 
%the corresponding event number appearing in the right relative proportion. 
When running in this mode, the user must specify which processes 
(and corresponding channels) should be considered in the event generation.
One can choose to produce events for all possible processes, or just
for a specified subset of them. The simplest option is generating
events for an individual process. 
%This would allow to study in detail some specific reaction, and its 
%peculiarities.
In any case, for a meaningful generation, the {\it grid-files} of all channels 
corresponding to each selected process must be included in {\tt r.in}. From 
the {\it grid-files}, {\tt ionesh=1} mode reads all necessary informations 
for the {\it hit-or-miss} selection.
\par\noindent
Phase-space grids and integrand maxima are prepared according to the cuts
specified during the {\tt ionesh=0} pre-run. When running in {\it one-shot}, 
one can impose new kinematical cuts. This option is implemented as follows. 
%As regards to kinematical cuts, the {\it one-shot} mode provides additional
%options which are implemented as follows. 
The structure of the common inputs, given in \refse{sec:singleprocess}, is 
exactly repeated in the cut section specific of the {\it one-shot} mode. The 
corresponding variables 
are the same as those in the common input section; they are just renamed with 
a suffix  - {\tt os} - appended. These additional cuts, operating at 
generation level, are obviously effective only if more restrictive than the 
common ones. The reason for doubling the cuts is the following.
There are different attitudes concerning signal selection and cuts.
One possible choice is to generate unweighted events with the loosest 
conceivable setup, 
and apply cuts directly on the produced event sample. This gives more 
freedom in varying the setup, according to the analysis at hand, without 
redoing the event generation; the drawback is an overproduction of events in 
regions which might not be of any interest. In \phanosp, this strategy 
traslates in producing both phase-space grids and generated events with the 
same setup. 
%The user should fill in the {\it input-file} with wide cuts, and keep them 
%unaltered when passing from {\tt ionesh=0} to {\tt ionesh=1}. 
A different general attitude is to implement cuts during event generation, in 
order to produce a sample already focused on the particular analysis to be 
performed. In this case, \pha provides two options. One can choose to run in 
{\it single-process} mode under the 
preferred cuts, in order to prepare grids specific for a certain study, and 
generate events with those same grids and cuts. This case does not differ 
from the previous one, as to the {\it input-file}. Otherwise, one could also 
produce phase-space grids with a looser setup (to retain all possible 
information) and impose more restrictive cuts later, when running in 
{\it one-shot} mode. Common cuts in {\tt r.in} must be kept 
identical in both runs. What changes is the set of cuts specific 
of the {\it one-shot} mode. Let us notice that, in this latter case, 
phase-space grids are prepared once for all, and one can perform different 
analyses by simply varying the {\it one-shot} specific cuts.
%Phase-space grids and integrand maxima are prepared according to the cuts
%specified during the {\tt ionesh=0} pre-run. When running in {\it one-shot}, 
%one can impose new kinematical cuts, but only if they are more restrictive 
%than the cuts used for preparing the grids.
\par\noindent
After choosing the processes to be represented in the event sample, one should 
specify the desired number of unweighted events. Once produced, these events 
are recorded in the file {\tt phamom.dat}, using the information stored in the 
two {\tt COMMON BLOCK}, {\tt HEPRUP} and {\tt HEPEUP}, according to the \LHP.
If required, each generated event is then passed to {\tt PYTHIA} 
for showering and hadronization via a call to {\tt PYEVNT}.

\section{Sample results}
\label{sec:sample}

In this section we present some applications of \phanosp. In particular, we 
focus on the Higgs signal and its electroweak irreducible background. In the 
class of processes we are addressing in this first version of the code, Higgs 
production is dominated by vector boson fusion followed by the Higgs decay 
into $\PW\PW$ pairs. This gives rise to a well known distinctive signature 
with two forward/backward tagging jets, and two quarks and one charged lepton 
from the $\PW$'s in the central region. In the following, we show examples of 
cross sections and distributions for two values of the Higgs mass: $M_\PH$=140 
and $M_\PH$=500 GeV. In the first case, the Higgs width is extremely narrow
and $\PW\PW$ pairs are produced below threshold. In the latter case, the Higgs 
resonance is rather broad, and the $\PW$'s are generated around and above 
their on-shell values. 
\par\noindent
After producing phase-space grids, we have generated two samples of one 
million unweighted events each, for all possible processes with one muon in 
the final state: $\Pp\Pp\to 4\Pq +\mu\nu_\mu$. In our notation $\mu\nu_\mu$ 
indicates both $\mu^-\bar\nu_\mu$ and $\mu^+\nu_\mu$. The produced event 
samples, one for each Higgs mass, are thus representative of all reactions 
shown in Table 1, including all possible quark flavours. 
%We have moreover included both charge conjugate and family related processes. 
%This implies that we
We consider a total of 644 processes. Not all of them contain
the Higgs signal. Some channels only contribute to the irreducible background,
but they must be included for any meaningful analysis.      
\noindent
%We work under the following setup. 
For the Standard Model parameters we use the input values:
\bq
\begin{array}[b]{lcllcllcl}
\MW & = & 80.40~\GeV, \qquad &
\MZ & = & 91.187~\GeV, \\
\Mt & = & 175.0~\GeV, &
\Mb & = & 4.8~\GeV, \\
\Gamma_\PW & = & 2.042774~\GeV, &
\Gamma_\PZ & = & 2.5007~\GeV.
\end{array}
\label{eq:SMpar}
\eq
We adopt the so called $G_{\mu}$-scheme (see Appendix A), and use the CTEQ5L 
parton distributions \cite{cteq} at the factorization scale:
\bq
Q^2={1\over 6}\sum_{i=1}^6\PT^2(i) 
\eq
where $\PT (i)$ is the transverse momentum of the {\it i}-th final state 
particle. We have implemented a general set of cuts, appropriate for LHC 
analyses. For the charged lepton we require
\bq
\PE (l)>20~\GeV~~~~~~~~\PT(l)>10~\GeV~~~~~~~~|\eta_l|< 3
\label{eq:cut1}
\eq
%where $\PE$ represents the energy, $\PT$ the transverse momentum, and 
%$\eta$ the pseudo-rapidity defined as 
%$\eta=-\log\left (\tan(\theta /2)\right )$, $\theta$ being the
%polar angle of the particle with respect to the beam.
Analogously, for the quarks we have
\bq
\PE (j)>20~\GeV~~~~~~~~\PT(j)>10~\GeV~~~~~~~~|\eta_j|< 6.5~~~~~~~~
m_{jj^\prime}> 20~\GeV
\label{eq:cut2}
\eq
\unitlength 1cm
\begin{picture}(15.,8.5)
\put(-0.3,2.){\includegraphics{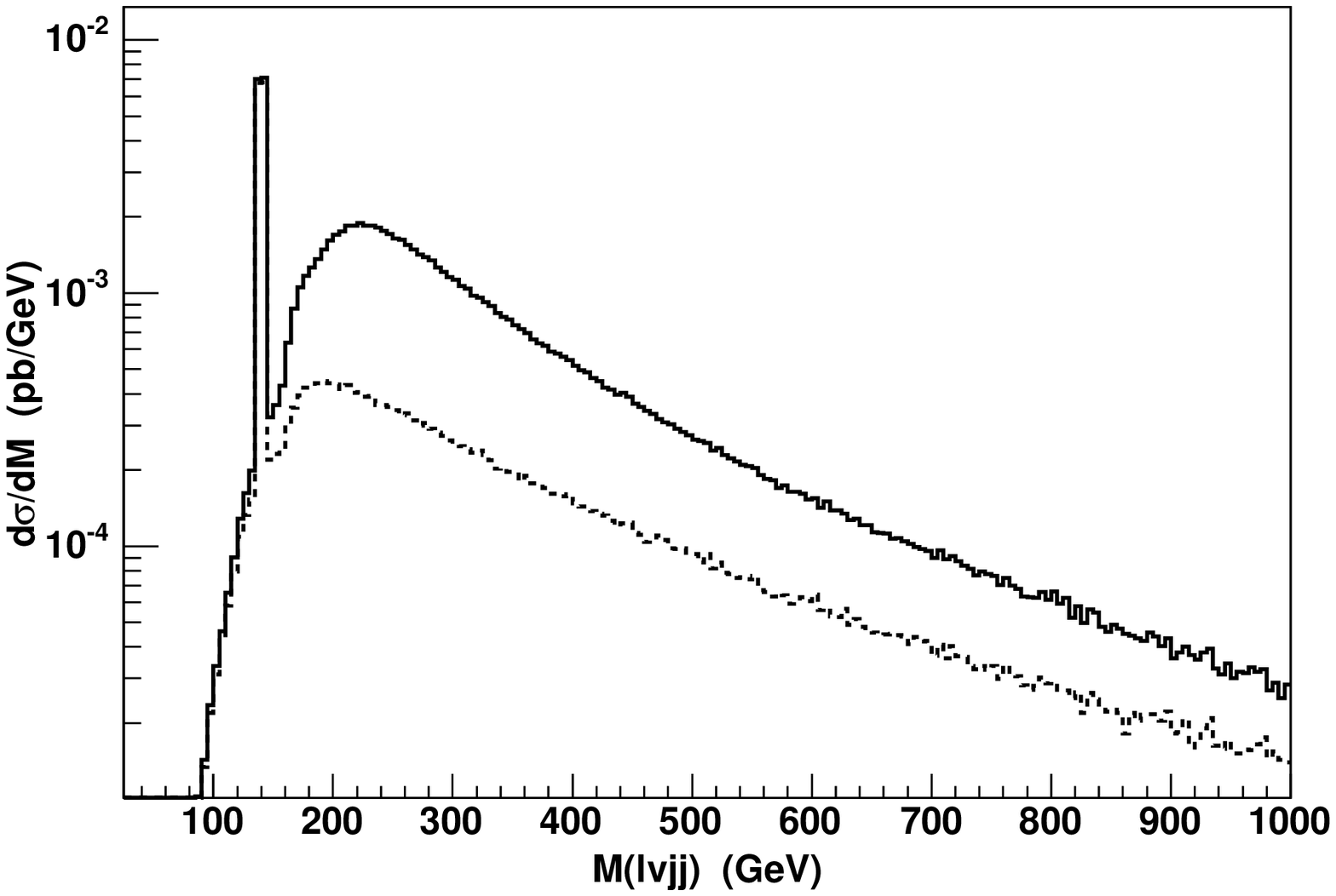}}
\put(7.2,2.){\includegraphics{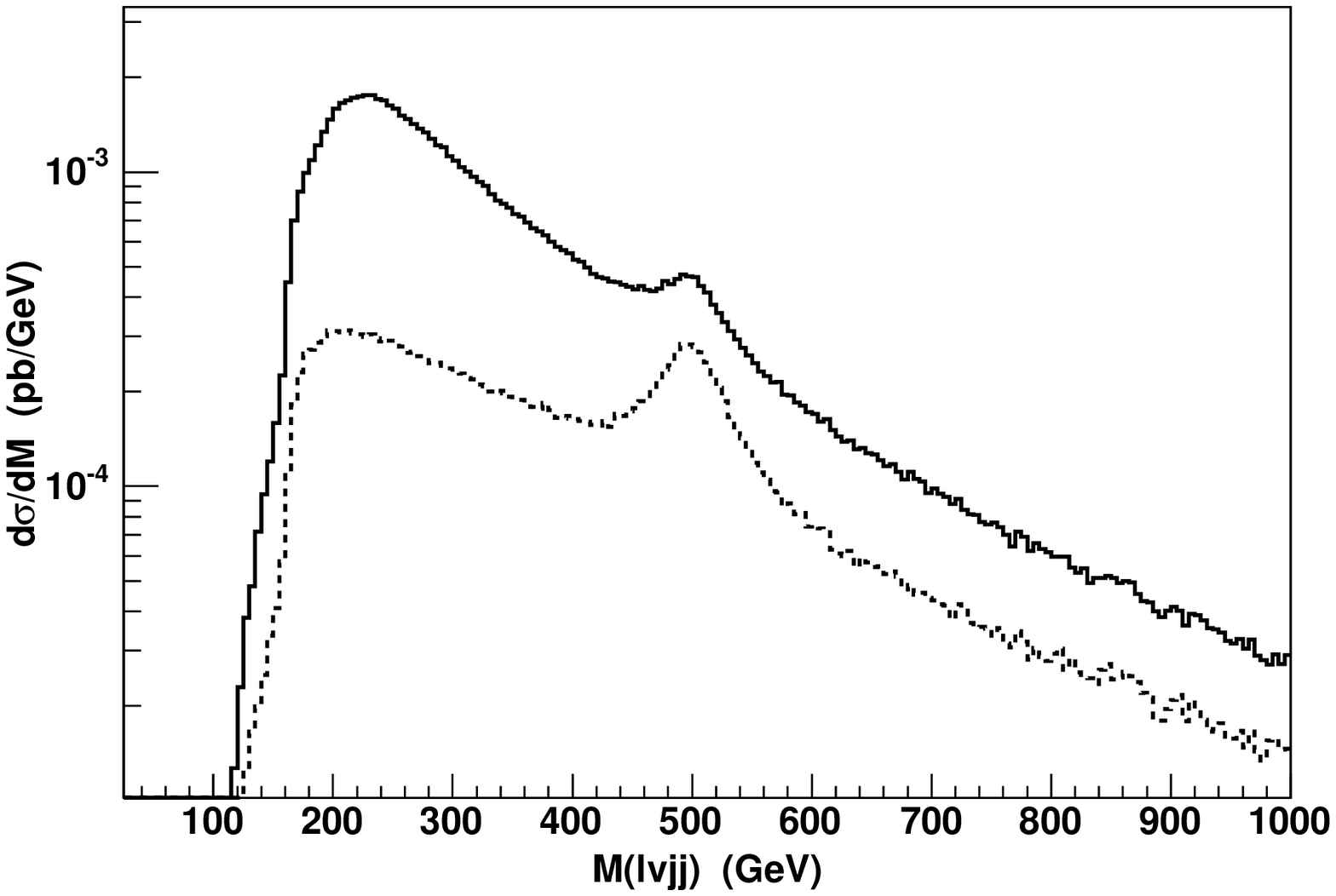}}
\put(-0.3,-4.5){\includegraphics{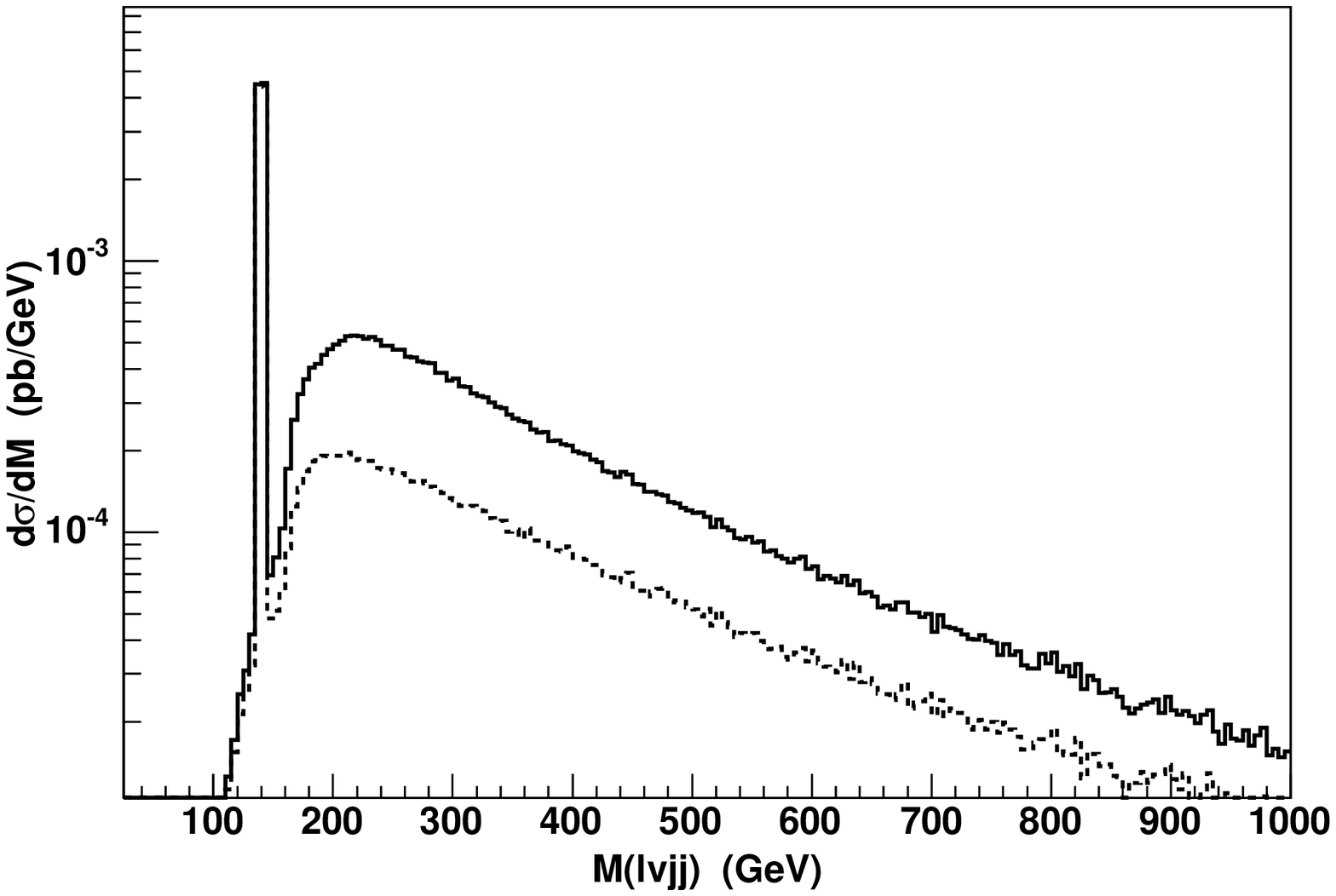}}
\put(7.2,-4.5){\includegraphics{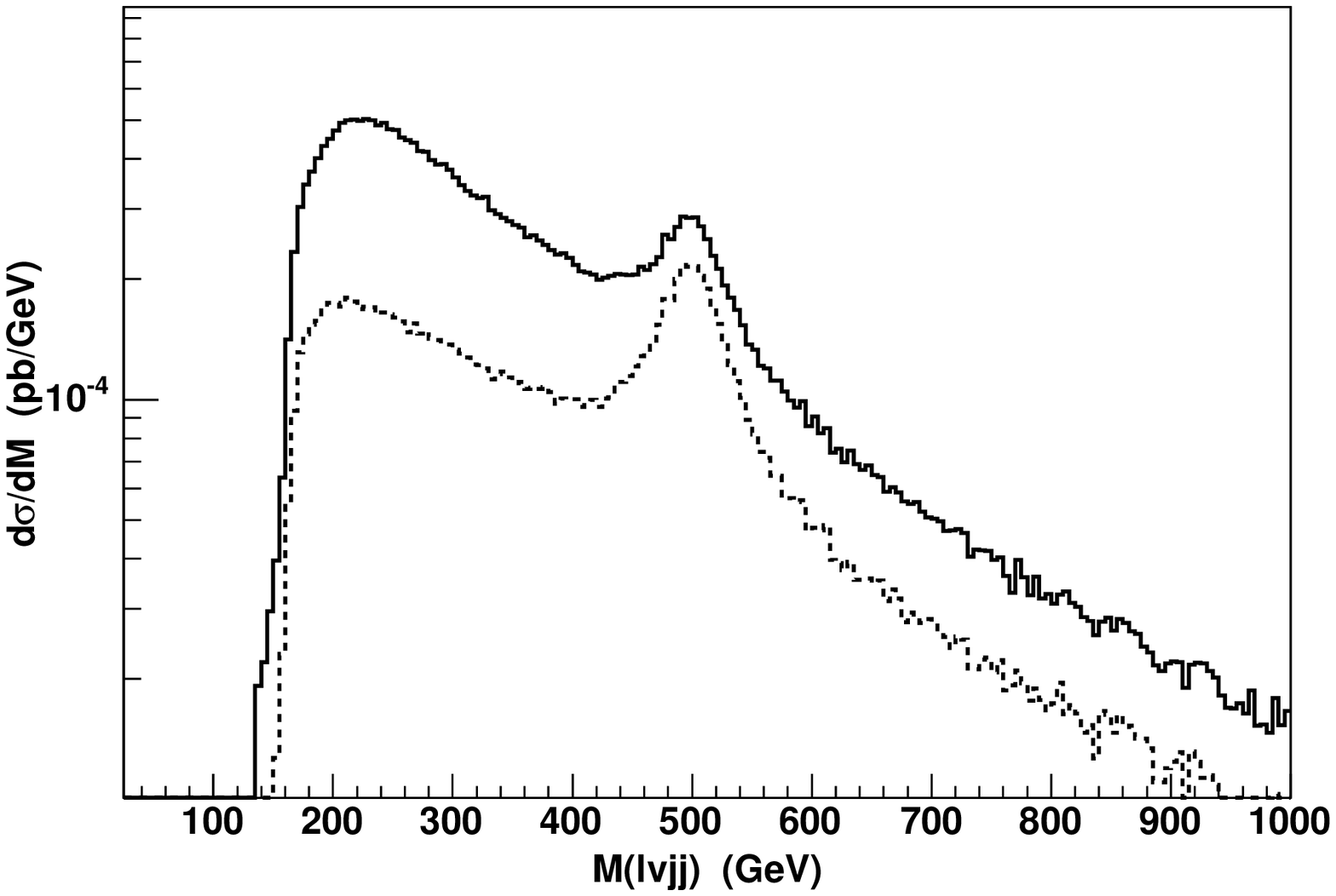}}
\end{picture}
\vskip 5.cm
Figure 3: Invariant mass distribution of the two leptons and the two central 
jets for 

$M_\PH$=140 and $M_\PH$=500 GeV, left- and right-hand side respectively. The 
solid curve in-

cludes all processes, the dashed one only final states with no $\Pb$-quark. 
The two upper 

plots have basic acceptance cuts, the lower ones include additional
forward-backward 

jet requirements, as explained in the text.
\vskip 1.5cm
\noindent
where $m_{jj^\prime}$ denotes the invariant mass of any jet pair. These cuts 
approximately simulate detector acceptance, and are common to all results 
presented in the following. 
In order to analyse the Higgs signal, we have plotted in Fig. 3 the total 
invariant mass of the two most central quarks, the muon and the neutrino,
which are supposed to originate from the Higgs decay into $\PW\PW$. 
The two upper plots include the basic acceptance cuts of eqs.
\refeq{eq:cut1}-\refeq{eq:cut2}. In the lower ones, we have also specifically
required two forward and backward jets, two central quarks, and one central 
charged lepton as follows:
\bq
1 <|\eta_{jf}|< 6.5 ~~~~~~~~-6.5 <|\eta_{jb}|< -1~~~~~~~~|\eta_{jc}|< 3~~~~~~~~
|\eta_l|< 3
\label{eq:cut3}
\eq  
\begin{table*}[htb]
\label{table:2}
\newcommand{\m}{\hphantom{$-$}}
\newcommand{\cc}[1]{\multicolumn{1}{c}{#1}}
\renewcommand{\tabcolsep}{0.7pc} % enlarge column spacing
\renewcommand{\arraystretch}{1.2} % enlarge line spacing
\begin{tabular}{@{}llllll}
\hline
$\PM_\PH$ (GeV) & $\sigma_{0b}$ (pb) & $\sigma_{1b}$ (pb) & 
$\sigma_{2b}$ (pb) & $\sigma_{3b}$ (pb) & $\sigma_{4b}$ (pb) \\
\hline
140 & 0.19844(4) & 0.11174(2) & 0.14502(4) & 0.6290(4)$\times 10^{-2}$ &
0.1395(3)$\times 10^{-3}$\\
\hline
500 & 0.12982(2) & 0.10767(2) & 0.14266(3) & 0.5507(4)$\times 10^{-2}$ &
0.956(3)$\times 10^{-4}$\\
\hline
\end{tabular}\\[2pt]
\caption{Cross sections for processes with $n$ $\Pb$-quarks/antiquarks in the 
final state, with $n=0,4$, for two values of the Higgs mass. Basic acceptance
cuts are included.}
\end{table*}
\par\noindent
where $jf$, $jb$ and $jc$ indicate forward, backward and central jets 
respectively. From the figure, one can clearly see the 
Higgs peak and the huge continuum background. 
A realistic study with detector simulation and reconstruction effects is 
needed to determine the actual shape of the peak, and the signal to background 
ratio. In this simple illustration, to clarify  the origin of the background, 
we have considered two sets of events for each Higgs mass and 
each setup. One is the full sample, containing events from all available 
processes. The other represents a subset, where only final states with no 
$\Pb$-quarks are included. The Higgs signal is essentially the same in the two 
cases, which instead differ substantially outside the resonance.
%The two upper plots in fig. 3 show that the two sets almost coincide at the 
%Higgs peak, while they differ substantially outside the resonance. 
The latter sample, which is much cleaner 
and does not suffer from possible electroweak top background, includes about 
one half(third) of the total number of events for $M_\PH$=140(500) GeV.
If one only imposes basic acceptance cuts, final states with $\Pb$-quarks are 
dominant. In order to see this in more details, we show in Table 2 cross 
sections for all processes with $n$ outgoing $\Pb$-quarks, where 
$0\le n\le 4$. Most of the contribution comes from processes with one 
and two b's in the final state, which in our sample are dominated by
electroweak single-top and $\Pt\bar\Pt$ production, respectively. 
If we require two forward-backward jets as in eq.\refeq{eq:cut3}, the signal 
to background ratio improves even in presence of $\Pb$-quarks in the final 
state, as shown in the two lower plots of Fig. 3. This suggests that the 
possibly dangerous top background to the Higgs search can be reduced either 
employing b-tagging techniques or imposing appropriate cuts. This analysis 
goes beyond the scope of this paper,
in which we simply present the potentiality of \pha for phenomenological 
studies.

%\unitlength 1cm
%\begin{picture}(15.,8.5)
%\put(-1.,2.){\special{psfile=mh140_etacut.eps hscale=40 vscale=40 angle=0}}
%\put(6.5,2.){\special{psfile=mh500_etacut.eps hscale=40 vscale=40 angle=0}}
%\end{picture}
%\vskip -1.5cm

\section{Conclusions}
\label{sec:conclusions}

The analysis of six-fermion final states is an important task at the LHC, 
owing to the several interesting subprocesses involved. These include 
Higgs and top production, vector boson scattering, and triple gauge boson
production. In this paper, we have presented the Monte Carlo event generator 
\phanosp, which in this first version computes all processes
$\Pp\Pp\to 4\Pq +\Pl\nu_{\Pl}$ at \cal{O}($\alpha^6$). 
\pha works with exact matrix elements. It employs a new {\it iterative-adaptive
multichannel} method for the phase-space integration. 
%The definition of mappings and channels is automatically performed. 
The algorithm considerably improves integral convergence and generation 
efficiency. The code makes use of the {\it one-shot} technique, which allows 
the user to generate in a single run an event sample fully representative of 
all available
final states (of the order of 1000). Upon request, the unweighted parton-level 
events are passed to hadronization packages via the \LHP, and eventually to
detector simulation codes. In this way, \pha can provide realistic event
samples, merging complete and precise theoretical computations for 
six-fermion processes with a detailed simulation of the experimental 
apparatus.   
\par\noindent
We have discussed in detail the general features of \phanosp. Some examples of 
the performance of the code have been shown. In particular, we have presented 
cross sections and distributions relevant to Higgs production, including all 
final states with one muon, $\Pp\Pp\to 4\Pq +\mu\nu_\mu$. 
The flexibility of the underlying concepts and the general structure
of \pha makes it easy to accomodate future developments. Enabling the code to 
calculate all processes $\Pp\Pp\to 6\Pf$ at \cal{O}($\alpha_s^2\alpha^4$) is 
the most important evolution planned for the near future. 

\vskip 0.3cm
\noindent
The first version of the program, {\tt PHASE~1.0}, can be downloaded from the 
following {\tt URL}: 
{\tt http://www.to.infn.it/$^\sim$ballestr/phase/}. 
All new versions of the code will be posted in this website.

\section*{Acknowledgements}

We thank Chiara Mariotti for the constant interest in our work and for
stimulating discussions and suggestions. Giuseppe Bevilacqua and Sara 
Bolognesi are gratefully acknowledged for their advice and tests.
Fabio Maltoni is acknowledged for comparisons. All diagrams have been drawn 
with JaxoDraw \cite{jaxodraw}. 

\begin{appendix}

\section{Parameters}
\label{sec:parameters}

Standard model parameters are defined in the routine {\tt coupling.f}.
%Vector boson and fermion masses are fixed in a {\tt DATA} statement,
%and may be changed. 
In our notation, {\tt rmw}, {\tt rmz}, {\tt rmt}, and {\tt rmb} are the 
$\PW$, $\PZ$, top and bottom 
masses respectively. The total $\PW$ and $\PZ$ widths are given by
{\tt gamw} and {\tt gamz}. Higgs and top widths are computed in the same
routine by standard formulas. 
\par\noindent
As a default, \pha employs the $G_\mu$-scheme
defined by the input set: $M_\PW$, $M_\PZ$ and $G_F$. According to this 
scheme, the calculated parameters are   
\beanon
sin^2\theta_\PW = 1-(M_\PW/M_\PZ )^2~~~~~~~~
\alpha_{em}(M_\PW )={\sqrt{2}\over{\pi}}G_FM^2_\PW sin^2\theta_\PW
\eeanon
\noindent
where $\theta_\PW$ is the weak mixing angle, and $\alpha_{em}$ the 
electromagnetic fine structure constant. 
\par\noindent
The code uses the {\tt CTEQ5*} Pdf parametrization, where the {\large *} 
indicates 
the possible schemes. As a default, we have implemented the {\tt LO-}scheme.
This can be modified by the user through the variable {\tt Iset} defined in 
the main body of the program, {\tt phase.f}.

\section{Input-file}
\label{sec:input}

In the following sections, 
%we explain how to fill in the two {\it input-files} for the two \pha running 
%modes, {\it single-process} and {\it one-shot}.
we describe how to use input parameters and flags to exploit the various 
possibilities of \phanosp. 
%Before discussing the basic informations the code 
%needs to get started, let us comment on the format of the input itself. 
The syntax of the input is almost identical to the one required by the CERN 
library routine {\tt FFREAD}. Routines internal to \pha are however used 
({\tt iread, rread}), so that real variables can (and must) be given in double 
precision. All lines in the {\it input-file} must not exceed 80 characters.
Writing {\tt \large *} or {\tt \large C} characters at the beginning of a 
line identifies it as a comment line. Comment lines can be freely interspersed 
within the {\it input-file}, with the only obvious exception that they must 
not interrupt a list of input values for a single array variable. The name of 
the variable to be read must be specified as the first word of a line. Its 
value (values) must follow it. The list of 
values can span several lines. A practical feature of the input routine is 
that variables, which are not needed to be specified in a given run, can be 
left in the {\it input-file} without any harm. They are simply ignored. 
The input values which are actually read are then written in the 
{\it output-file}. Two sample {\it input-files} for \pha are provided in the 
program package,
{\tt inp.st0} for {\tt ionesh=0} and {\tt inp.st1} for {\tt ionesh=1}.
When running the program, they have to be renamed. The actual {\it input-file} 
must always be called {\tt r.in}. All energy values must be given in GeV, 
while angles are in degrees. Kinematical variables are all defined in the 
collider frame. For yes/no flags, we adopt the convention that {\tt 0} 
corresponds to {\tt no} and {\tt 1} to {\tt yes}.
All entries in the following sections, which describe input variables in more
detail, are in the form:

\inpe{variable-name}
\vsk
or

\inp{variable-name}{full-list-of-possible-values: val1/val2/\dots}
\vsk

\subsection{Common inputs}
\label{sec:commoninput}

\inp{ionesh}{0/1} this flags selects the basic operation mode of \pha
as explained in \refse{sec:modes}.

\inpe{idum} random number generation seed. Must be a large 
negative integer.

\inpe{ecoll} total center of mass energy of $\Pp\Pp$ collisions.

\inpe{rmh} Higgs mass. Setting {\tt rmh} to a negative number allows the
user to switch off Higgs diagrams.

\inp{i\_ccfam}{0/1} if {\tt i\_ccfam=0} only processes explicitely required by 
the user are computed. If {\tt i\_ccfam=1} the required processes are computed 
along with the reactions obtained interchanging first and second family of 
quarks and antiquarks, and with the reactions obtained by charge conjugation. 
For instance, if the user-specified process is
\begin{equation}
u\bar{u}\rightarrow b\bar{b} c\bar{s}\mu^-\bar\nu_\mu
\end{equation}
with {\tt i\_ccfam=1}, all the following processes are computed or generated
in the same run:
\beanon
    u \bar{u} \rightarrow b \bar{b} c \bar{s} \mu^- \nu_\mu~~~~~~
    c \bar{c} \rightarrow b \bar{b} u \bar{d} \mu^-\nu_\mu\\
    \bar{u} u \rightarrow \bar{b} b \bar{c} s \mu^+ \bar{\nu_\mu}~~~~~~
    \bar{c} c \rightarrow \bar{b} b \bar{u} d \mu^+ \bar{\nu_\mu}\\
\eeanon
\noindent
This computation involves a sum over Parton Distribution Functions, and it 
gives different cross section and different grids, if compared to the 
{\tt i\_ccfam=0} case. When generating events in ({\tt ionesh=1}) mode, it is
thus important to give {\tt i\_ccfam} the same value used for preparing the
{\it grid-files}.

\subsubsection{Cuts}
All cuts in \pha are meant at parton level, before showering and hadronization.
As described in \refse{sec:singleprocess}, two types of predetermined cuts are
provided in \phanosp. The basic one simulates detector acceptance and 
separation
criteria. The corresponding variables are characterized by the suffixes
{\tt \large lep} and {\tt \large j}, which refer to {\it charged} lepton 
and quark/antiquark, respectively. 
%The suffix {\tt \large lep} refers to {\it charged} leptons, while 
%{\tt \large j} refers to quarks/antiquarks. 
The second kind of cuts is instead focused on Higgs search and vector boson 
scattering analyses. In particular, we define the most forward and most 
backward jets. The remaining two jets are called central. In this case, the 
suffixes {\tt \large jf}, {\tt \large jb} and {\tt \large jc} denote forward, 
backward and central jets, respectively. A yes/no flag specifies whether 
the corresponding cut is activated or not. The name of this flag in most cases 
is the name of the corresponding variable with  {\tt i\_} prepended. 
Exceptions to this rules will be pointed out; in all other cases we will give 
only the variable name and the corresponding flag will be understood.
Variables are defined as follows:

\inpe{e\_min\_lep} minimum energy of charged leptons.

\inpe{pt\_min\_lep} minimum transverse momentum of charged leptons.

\inpe{eta\_max\_lep} maximum absolute value of charged lepton 
pseudo--rapidity.

\inpe{ptmiss\_min} minimum missing transverse momentum (at present it coincides 
with the neutrino transverse momentum).

\inpe{e\_min\_j} minimum energy of quarks/antiquarks.

\inpe{pt\_min\_j} minimum transverse momentum of quarks/antiquarks.

\inpe{eta\_max\_j} maximum absolute value of quark/antiquark pseudo--rapidity.

\inp{i\_eta\_jf\_jb\_jc}{0/1} specifies whether the following triplet of cuts 
are activated:
\begin{flushright}
\begin{minipage}{.95\textwidth}
\inpe{eta\_def\_jf\_min} minimum value of the pseudo--rapidity of the most 
forward quark/antiquark.

\inpe{eta\_def\_jb\_max} maximum value of the pseudo--rapidity of the most
backward quark/antiquark.

\inpe{eta\_def\_jc\_max} maximum absolute value of the pseudo--rapidity
of the remaining two ({\it central}) quarks/antiquarks.
\end{minipage}
\end{flushright}

\inpe{pt\_min\_jcjc} minimum total transverse momentum of the two central 
quarks/antiquarks.

\inpe{rm\_min\_jj} minimum invariant mass of quark/antiquark pairs.

\inpe{rm\_min\_jlep} minimum invariant mass of any pair of charged--lepton and 
quark/antiquark.

\inpe{rm\_min\_jcjc} minimum invariant mass of the two central 
quarks/antiquarks.

\inpe{rm\_max\_jcjc} maximum invariant mass of the two central 
quarks/antiquarks.

\inpe{rm\_min\_jfjb} minimum invariant mass of the most forward and 
most backward quark/antiquark.

\inpe{eta\_min\_jfjb} minimum absolute value of the difference in 
pseudo--rapidity between most forward and most backward quark/antiquark.

\inpe{d\_ar\_jj} minimum separation in 
$\Delta R = \sqrt{\Delta \phi +\Delta \eta}\ $  between any two 
quarks/antiquarks.

\inpe{d\_ar\_jlep} minimum separation in $\Delta R$ between any quark/antiquark
and charged lepton.

\inpe{thetamin\_jj} minimum angular separation between two quarks/antiquarks.

\inpe{thetamin\_jlep} minimum angular separation between quarks/antiquarks and 
charged leptons.

\inp{i\_usercuts}{0/1} determines whether additional user-specified cuts are 
required. These requirements must be implemented in a routine called 
{\tt IUSERFUNC}, an example of which is provided in the program package.

\subsection{{\tt ionesh=0} input}

\inpe{iproc} specifies the desired process using the standard Monte Carlo 
particle numbering scheme:

\begin{table}[hbt]\centering
\begin{tabular}{|c|c|c|c|c|c|c|c|c|c|c|}
\hline
 $d$ & $u$ & $s$ & $c$ & $b$ & $e^-$ & $\nu_e$ & $\mu$ & $\nu_\mu$ &
  $\tau$ & $\nu_\tau$\\
\hline
 $1$ & $2$ & $3$ & $4$ & $5$ & $11$ & $12$ & $13$ & $14$ &
  $15$ & $16$\\
\hline
\end {tabular}
\end {table}
\noindent
Antiparticles are coded with the opposite sign. The variable {\tt iproc} is
an eight-component vector, where the first two entries represent the initial
state partons. As an example, {\tt iproc=(3,-4,2,-2,3,-3,13,-14)} corresponds
to the reaction $s\bar c\rightarrow u\bar u s\bar s \mu\nu_\mu$. In 
{\tt ionesh=0} mode, the process is computed exactly as written by the user,
assuming the first incoming particle to be moving in the $+z$ direction and 
the second one in the $-z$ direction (the realistic case at a $pp$ collider,
which accounts for the exchange of the two initial particles, is implemented
only in {\tt ionesh=1} mode with the flag {\tt i\_exchincoming}). 
%In this case the program proceeds in two steps. The first one is called 
%thermalization. It determines the relative weight of each channel in the 
%multichannel integration and it produces a first instance of phase space 
%grids, one per channel. These grids are then used as a starting point for the 
%second step which consists of one integration per channel. Each integration 
%will typically consist of several iterations and at each iteration the phase 
%space grid will be refined in an effort to decrease the overall variance. A 
%number of iterations between 3 and 5 is normally the best choice. If higher 
%precision is requested it is usually more convenient to increase {\tt ncall} 
%rather than {\tt itmx}. The user must be aware of the fact that if no point 
%survives the cuts during an iteration, either during thermalization or at the 
%integration stage, \veg will stop with an error.
%When {\tt ionesh=0}, the process is computed exactly as specified by the user,
%assuming the first incoming particle to be moving in the $+z$ direction and 
%the second one in the $-z$ direction. Even though unrealistic at a
%$pp$ collider this can be useful for testing and for specialized studies.
%This behaviour can be modified when {\tt ionesh=1} by the flag
%{\tt i\_exchincoming}. As a consequence cross sections computed with 
%{\tt ionesh=1} can be different from those computed with {\tt ionesh=0}.

\inpe{acc\_therm} integration accuracy in thermalization. When this accuracy 
is reached for a given channel, thermalization of that channel stops.

\inpe{ncall\_therm} maximum number of points for each iteration during 
thermalization.

\inpe{itmx\_therm} maximun number of iterations used to evaluate each integral 
in thermalization. 

\inpe{acc} accuracy of the actual integration. When this accuracy is reached 
for a given channel, integration of that channel stops.

\inpe{ncall} maximum number of points for each iteration of the actual 
integration. In general, {\tt VEGAS} uses a number of {\tt ncall\_therm} and 
{\tt ncall} lower than the input ones. The actual value is written in the 
{\it output-file}, where also the number of points which survive all the cuts 
({\tt effective ncall}) is reported. 

\inpe{itmx} maximum number of iterations used to evaluate the integral and 
refine the grid. A number of iterations between 3 and 5 is normally the best 
choice. If higher precision is requested, it is usually more convenient to
increase {\tt ncall} rather than {\tt itmx}. The user must be aware of the fact
that if no point survives the cuts during an iteration, either during 
thermalization or at the integration stage, {\tt VEGAS} will stop with an 
error.

\inp{iflat}{0/1} this yes/no flag must be set to {\tt 1} in order to produce 
the phase-space grids for later unweighted event generation. If {\tt iflat=1},
the program also returns the maximum of the integrand function $w0$ produced 
in the next--to--last iteration, the maximum $w1$ produced in the last 
iteration, and the number of points with weight greater than $w0*scalemax0$ 
visited during the last iteration. By default we take $scalemax0=1.1$.
The maximum $w1$, stored in
{\tt PHAVEGAS0i.DAT}, is then used in {\it one-shot} mode, which employs the
{\it hit-or-miss} method for the unweighted event generation.
%If {\tt iflat=1} the following parameter must be set:
%\begin{flushright}
%\begin{minipage}{.95\textwidth}
%\inpe{scalemax0} factor by which the maximum value of the differential cross 
%section is multiplied for the {\it hit-or-miss} selection. This coefficient 
%can be used to compensate for the fact that the maximun determined by the 
%program may be smaller than the true maximum. Setting this parameter too high 
%would decrease the efficiency in generation.
%\end{minipage}
%\end{flushright}

\subsection{{\tt ionesh=1} input}

\inpe{nunwevts} number of unweighted events the user desires to produce. The 
program stops only when this number has been reached.

\inpe{scalemax} factor used to replace the integrand maximum in the 
{\it hit-or-miss} procedure.  
This coefficient multiplies the maximum value of the differential cross 
section, found during the phase-space grid preparation. It can be used to 
compensate for the fact that the maximum determined by the program may be 
smaller than the true maximum. Setting this parameter too high would decrease 
the efficiency in generation. On the other side, it is not advisable to 
lower the maximum value for a more efficient unweighting. The generated
sample could be biased.

\inp{iwrite\_event}{0/1} yes/no flag which decides whether the generated 
events are recorded in the file named {\tt phamom.dat}. For writing this file,
the code uses the information stored in the two {\tt COMMON BLOCK} 
{\tt HEPRUP} and {\tt HEPEUP} according to the \LHP.

\inp{ihadronize}{0/1} yes/no hadronization required. If {\tt ihadronize=1}, 
each generated event is passed to {\tt PYTHIA} for showering and 
hadronization via a call to {\tt PYEVNT}, using the \LHP.

\inp{i\_exchincoming}{0/1} yes/no flag which symmetrizes the initial state.
If the value is set to zero, the process is generated exactly as required,
assuming the first incoming particle to be moving in the $+z$ direction and 
the second one in the  $-z$ direction. Otherwise, the two initial particles
are assigned at random to the two protons, doubling the corresponding cross 
section if they are not identical. As a consequence, cross sections computed 
in {\tt ionesh=1} mode can be different from those computed in {\tt ionesh=0}.

\inp{i\_emutau}{0/1/2} determines which charged leptons are present in the 
generated sample. If {\tt i\_emutau=0}, only events containing the charged
lepton specified by the user in the vector {\tt iproc} will be generated.
If {\tt i\_emutau=1}, events containing  $\mu$ and events containing  $e$
will be generated with the same frequency. Finally, if {\tt i\_emutau=2} 
events containing  $\mu$, events containing  $e$ and events containing 
$\tau$ will be generated with the same frequency.
%The flag {\tt i\_ccfam}, specified in the common part of the input, 
%determines whether or not the corresponding positively charged leptons are 
%produced.

\subsubsection{Cuts}

\inpe{iextracuts} determines whether additional cuts are required at the 
generation stage. The {\it input-file} for {\tt ionesh=1} must contain the 
same set of cuts used for generating phase-space grids, and defined in 
\refse{sec:commoninput}. As a consequence, additional cuts will be effective
only if they are more stringent than those imposed in the {\tt ionesh=0} 
pre-run.
These extra cuts must be defined setting {\tt iextracuts=1}, followed by the 
new cut list. The names of the corresponding variables are equal to those in 
the common input section with the suffix - {\tt os} - appended.

\inpe{i\_usercutsos} determines whether additional user-specified cuts are 
required at the generation stage. These requirements must be implemented in a
routine called {\tt IUSERFUNCOS}, an example of which is provided in the code
package. Obviously, the comments concerning the relationship between cuts in 
the grid--production and event-generation stage also apply to the 
user-specified cuts.
 
\subsubsection{Processes}

\inpe{nfiles} number of {\it grid-files} ({\tt PHAVEGAS0i.DAT}) to be 
considered in generation. This input should be immediately followed, with no 
intervening blank line, by {\tt nfiles} filenames, each on a separate line, as 
in the following example:

\begin{verbatim}
nfiles   3
/home/user/dir1/phavegas01.dat
/home/user/dir1/phavegas02.dat
/home/user/dir2/phavegas01.dat
\end{verbatim}
\vsk
\noindent
All phase-space grids for each selected process must be included for a 
meaningful generation. In the example at hand, the first two files from the 
top represent the two {\it grid-files} of the two channels corresponding to 
the same process stored in directory {\tt dir1}. The last file contains the 
single channel grid of the process in directory {\tt dir2}.  

\end{appendix}

\end{document}